\pdfoutput=1
\documentclass{emulateapj}

\shorttitle{Variable stars in the VVV globular clusters I}
\shortauthors{Alonso-Garc\'ia et al.}

\begin{document}


\title{Variable stars in the VVV globular clusters. I. 2MASS-GC02 and Terzan~10}


\author{Javier Alonso-Garc\'{i}a\altaffilmark{1,2}, Istv\'an~D\'ek\'any\altaffilmark{2,1}, M\'{a}rcio Catelan\altaffilmark{1,2}, Rodrigo Contreras Ramos\altaffilmark{1,2}, Felipe Gran\altaffilmark{1,2}, P\'ia Amigo\altaffilmark{3,2}, Paul Leyton\altaffilmark{1,2}, and Dante Minniti\altaffilmark{4,1,2,5}}


\altaffiltext{1}{Instituto de Astrof\'{i}sica, Facultad de F\'isica, Pontificia Universidad Cat\'{o}lica de Chile, Av. Vicu\~na Mackenna 4860, 782-0436 Macul, Santiago, Chile} 
\altaffiltext{2}{Millennium Institute of Astrophysics, Av. Vicu\~{n}a Mackenna 4860, 782-0436 Macul, Santiago, Chile.}
\altaffiltext{3}{Instituto de F\'isica y Astronom\'ia, Universidad de Valpara\'iso, Av. Gran Breta\~na 1111, Playa Ancha, Casilla 5030, Chile}
\altaffiltext{4}{Departamento de Ciencias F\'isicas, Universidad Andr\'es Bello, Rep\'ublica 220, Santiago, Chile}
\altaffiltext{5}{Vatican Observatory, Vatican City State V-00120, Italy.}


\begin{abstract}
The VISTA Variables in the V\'ia L\'actea (VVV) ESO Public Survey is
opening a new window to study the inner Galactic globular clusters
using their variable stars. These globular clusters have been
neglected in the past due to the difficulties caused by the presence
of an elevated extinction and high field stellar densities in their
lines of sight. However, the discovery and study of any present
variables in these clusters, especially RR~Lyrae stars, can help to
greatly improve the accuracy of their physical parameters. It can also
help to shed some light on the interrogations brought by the
intriguing Oosterhoff dichotomy in the Galactic globular cluster
system. In a series of papers we plan to explore the variable stars in
the globular clusters falling inside the field of the VVV survey.  In
this first paper we search and study the variables present in two
highly-reddened, moderately metal-poor, faint, inner Galactic globular
clusters: 2MASS-GC02 and Terzan~10. We report the discovery of sizable
populations of RR Lyrae stars in both globular clusters. We use
near-infrared period-luminosity relations to determine the color
excess of each RR Lyrae star, from which we obtain both accurate
distances to the globular clusters and the ratios of the selective to
total extinction in their directions. We find the extinction towards
both clusters to be elevated, non-standard, and highly
differential. We also find both clusters to be closer to the Galactic
center than previously thought, with Terzan~10 being on the far side
of the Galactic bulge. Finally, we discuss their Oosterhoff
properties, and conclude that both clusters stand out from the
dichotomy followed by most Galactic globular clusters.
\end{abstract}





\section{Introduction}
\label{sec_intro}
Observations of most of the globular clusters (GCs) located in the
Galactic bulge region are severely affected by the effects of
reddening, both absolute and differential, and by high densities of
field stars \citep{val07,alo12}. This highly complicates the analysis
and interpretation of their color-magnitude diagrams (CMDs), and the
accuracy of the physical parameters obtained therefrom. This is
especially dramatic in the case of low-surface brightness,
inner-Galactic GCs, which complicates even their very detection
\citep[e.g.,][]{hur00,min11b,mon11}.

The analysis of any variable star content, especially RR~Lyrae, in the
inner Galactic GCs is invaluable for better determining their physical
parameters. RR~Lyrae are bright radial pulsators, present in
considerable numbers in most of the old Galactic GCs
\citep{cle01}. They have been shown to follow tight period-luminosity
relations in the near-infrared wavebands, especially in the $K$-band
\citep[e.g.,][]{lon86,lon90,cat04}, that can be used as a powerful
tool for distance determinations. But a systematic census and
measurement of accurate near-infrared light curves of RR~Lyrae
variables in the inner Galactic GCs have been pending until
now. Fortunately, the {\it VISTA Variables in the V\'ia L\'actea}
(VVV) ESO Public Survey provides a headway in improving this
situation. VVV is a multi-epoch, near-infrared survey that covers most
of the bulge region of our Galaxy, along with a region of the Southern
disk with high star formation, using ESO's 4-m VISTA Telescope in
Chile \citep{min10,sai11,cat11}. Numerous inner Galactic globular
clusters, 2MASS-GC02 and Terzan~10 among them, fall inside the area
covered by the VVV survey. This allows us to detect the variable stars
of these clusters, derive their periods and mean magnitudes, analyze
their light curves, determine their variability types, and employ
them, especially any discovered fundamental-mode RR~Lyrae pulsators,
for a more accurate determination of the physical parameters of their
host GCs.

A particular characteristic of the Galactic GCs is the Oosterhoff
dichotomy \citep{oos39,cat09a,smi11}. The GCs in the Milky Way are
clumped into two main groups: Oosterhoff I contains fundamental-mode
RR~Lyrae (RRab) with shorter periods ($\langle P_{\rm ab} \rangle \sim
0.55$~days), and Oosterhoff II have RRab's with longer periods
($\langle P_{\rm ab} \rangle \sim 0.64$~days), leaving an almost empty
gap, the so-called Oosterhoff gap, at $\langle P_{\rm ab} \rangle \sim
0.60\pm0.02$~days. This dichotomy seems not to be present in other
nearby extragalactic systems, which tend to be preferentially
Oosterhoff-intermediate.  The interpretation of the Oosterhoff
dichotomy in terms of the history of Galactic assembly has been
extensively debated. \citet{yoo02} theorized that the dichotomy is a
result of the decoupled age and metallicity distributions of the
Galactic GC system, in which the most metal-poor GCs are younger than
the oldest ''genuine'' Galactic ones and were accreted from
satellites. However, \citet{cat09a} argued that age determinations
from detailed case studies of halo Galactic GCs have not provided
enough evidence for this scenario. In any case, the Oosterhoff
dichotomy implies that the oldest components of the Milky Way cannot
have formed by pure accretion of the early counterparts of its
present-day dwarf galaxy satellites \citep{cat09a,cat09b}.  It is
important to add that our current sample of Galactic GCs with firm
knowledge about their Oosterhoff properties is still incomplete. In
particular, bulge GCs are largely unexplored in this respect, mostly
due to the lack of time series observations. Among the few exceptions
with known Oosterhoff properties, NGC 6441 and NGC 6388 stand out from
the Oosterhoff types of both Galactic and extragalactic GCs, with
$\langle P_{\rm ab} \rangle$ too long for their metallicities, forming
the so-called Oosterhoff III group
\citep{pri00,pri01,pri02,pri03,cor06}. This property has been
attributed to the presence of a helium-enhanced population of stars in
these clusters \citep{cal07,yoo08}, causing the RR Lyrae stars of this
population to have longer periods, although it remains unclear why
seemingly all the RR Lyrae stars are helium-enhanced in these two
clusters. By observing the yet unknown Oosterhoff properties of the
bulge GCs, we expect to gain new insights into the formation and
evolution of these clusters, and of the central components of the
Milky Way itself.

In a series of papers, we will explore the near-infrared time-domain
of the inner Galactic GCs in the framework the VVV survey,
characterize their variable star content, and derive the GCs' physical
parameters therefrom. In the present paper, the first of this series,
we have chosen to study two metal-intermediate, low-surface brightness
GCs that show some of the highest reddening in the VVV sample of GCs,
2MASS-GC02 and Terzan~10.

\section{Observations and data reduction}
\label{sec_obs}
Our observations are part of the VVV survey, that is being conducted
since early 2010 with the 4.1m VISTA Telescope in Cerro Paranal
Observatory, in Chile. The VIRCAM camera on the telescope consists of
16 detectors, each of them providing near-infrared images of
$11\farcm6\times11\farcm6$ with a pixel size of $0\farcs34$. We used
the stacks of 2 slightly dithered ($\approx20\arcsec$ in each
direction) individual exposures of the VVV fields that contain our
target GCs, 2MASS-GC02 and Terzan~10. The main characteristics of
these GCs are listed in Table~\ref{tab_gc}.  Since the field provided
by the stacked images is sparsely populated, with significant gaps
between the chips, a mosaic pattern is needed for a contiguous
coverage of every field observed in the VVV \citep{min10,sai12b}, and
every region covered in our study is observed a minimum of 2 times and
a maximum of 6 times per epoch. The individual images were reduced,
astrometrized and stacked by the Cambridge Astronomy Survey Unit
(CASU) using the VISTA Data Flow System (VDFS) pipeline
\citep{eme04,irw04,ham04}. Although so-called \emph{tiles},
i.e. mosaic images with contiguous coverage of the whole detector area
at every field, combined from 6 offset stacked images at each epoch,
are also provided by CASU, we decided to work with the individual
stacked images. This way we avoid problems in the photometry extraction related with combining offset images, which have different geometrical distortions for a given sky region producing highly space-varying and difficult-to-model point-spread functions (PSF) for the detected objects, and problems related with combining different chips, with significant differences in their dynamic range in some cases. A set
of 43 epochs in $K_s$ for 2MASS-GC02 and a set of 101 epochs in $K_s$
for Terzan~10 were available at the time of our variability study. We
also used the single-epoch $Z$, $Y$, $J$, and $H$ images for every
cluster to build the CMDs and calculate the stellar extinctions, as we
show in the following sections. Most epochs in $K_s$ correspond to
different nights, and the seeing in most of them was very good,
between $0.6\arcsec$ and $1.2\arcsec$. A summary of the observations
is provided in Table~\ref{tab_obs}.
 
PSF photometry was carried out on the individual processed stacked
images using an updated version of DoPHOT \citep{sch93,alo12}. CASU
also provides catalogs of aperture photometry, and we used them to
calibrate our PSF photometry into the VISTA system. We chose to
perform the PSF photometry out to a distance of $10\arcmin$ from the
cluster center, well beyond both GCs' tidal radii (see
Table~\ref{tab_gc}), to also explore a significant region around the
GCs, looking for extra-tidal variables and comparing cluster vs. field
ratios of variables in the surroundings of the GCs. Certainly this is
one of the advantages provided by the big area surveyed by
VVV. Photometry for the same object in the different available images
was cross-correlated using the STILTS package \citep{tay06}. The
generated light curves were next analyzed for variability.

\section{Variability analysis}
\label{sec_varana}
We started our variability analysis by performing iterative
$10{\sigma}$ and $5{\sigma}$ threshold rejections in the light curves,
in order to omit outliers that could heavily bias our analyses. Then,
we performed a first selection of candidate variable stars by taking
advantage of the correlated sampling of the time-series, i.e.,
multiple (2--6) points at each epoch, resulting from the subsequent
exposures with slightly offset pointings necessary to get a contiguous
coverage of the field, as described in Section~\ref{sec_obs}. These
batches of successive exposures are always taken within a few minutes
of time, which is by orders of magnitude shorter than both the time
between two separate batches, and the typical time-scale of stellar
variability of our interest. Therefore, the fluxes measured within
such a sequence can well be considered to sample the light curve at
the same epoch, thus the differential flux within them will be
dominated by noise. Likewise, the deviations of the fluxes measured
from the average brightness, for instance, will be expected to be well
correlated within such batches in case of intrinsically variable
stars, and rather uncorrelated otherwise. A well-known descriptive
statistic that was specifically designed to measure this kind of
correlations in astronomical time-series with such sampling is the
so-called Stetson's index \citep{wel93,ste96}. We employed the
generalized version of this statistic \citep[][Eq.~1]{ste96} for
pre-selecting a first set of variable star candidates, by requiring
the value of the Stetson index to be higher than a pre-determined
critical threshold.  The latter was set as the value of the index at
the $0.1\%$ significance level corresponding to pure Gaussian noise,
and was derived from Monte Carlo simulations for various different
numbers of observational epochs. This procedure resulted in our broad
{\em first selection} of candidate variable stars, including $\sim
10^4$ light curves in the areas surrounding both clusters.

Candidates pre-selected by their Stetson index were subjected to
frequency analysis. First, we binned the light curves in order to
simplify the assessment of spectral significance, i.e. we computed the
weighted averages of the magnitudes in each batch of stacked images,
resulting in light curves with one point per epoch. Then we computed
both the Generalized Lomb-Scargle periodogram \citep[GLS]{zec09} and
the phase dispersion spectrum \citep[PDM]{ste78} of each light curve,
and determined the primary frequency components from both methods in
the $[0-10]\,{\rm d}^{-1}$ frequency range, using a spectral
oversampling factor of 10. Significance levels of these components
were estimated analytically.  We selected the light curves that showed
periodic signals detected with better than $0.1\%$ significance by
both methods (not requiring the frequencies to coincide). We rejected
signals with close to integer ${\rm d}^{-1}$ frequencies in order to
exclude false alarms triggered by typical systematics of ground-based
observations \citep[see, e.g.,][]{kov05}.  However, we considered
light curves showing low-frequency spectral excess with a total
amplitude of $\geq0.1$~mag, which could be indicative of periodic
signals with periods longer than the time span of our observations, or
aperiodic variations such as transients. Our procedure resulted in our
narrow {\em second selection} of $\sim2000$ and $\sim5000$ variable
star candidates in the case of 2MASS-GC02 and Ter 10, respectively.

In the next step of our analysis, we performed a combined visual
inspection of both the phase diagrams and the images of each and every
object in the {\em second selection}, and rejected those which did not
meet some basic phenomenological criteria.  Most rejected light curves
had obvious signs of temporal saturation and/or blending and showed
false signals originating from these effects. In addition, a large
number of sources turned out to be fake variables, with periodic
signals emanating from the temporally contaminating flux from the
rotating diffraction spikes of nearby saturated stars (due to the
altazimuthal configuration of VISTA). As a result of this procedure,
we narrowed down our selection to 102 and 160 variables in the field
of 2MASS-GC02 and Ter~10, respectively.

As a final step in our time-series analysis, we performed an iterative
refinement procedure on the {\em unbinned} light curves. We performed
a non-linear Fourier fit using the frequencies from the GLS/PDM
analyses as initial values, visually optimizing the number of Fourier
orders for each object. In cases where the frequencies from the two
methods had a significant discrepancy, we adopted the one that
produced the smaller $\chi^2$. We then performed a threshold rejection
around the fit, and refitted the light curve until the procedure
converged in a final solution. The apparent $K_s$-band equilibrium
brightnesses of the stars were estimated by the intensity-averaged
magnitudes of the stars, computed from the Fourier fits to the
light curves. The total amplitudes of the light curves were computed
from the Fourier fits.

We tried to assign a variability type to the discovered variables, but
this task proved to be non-trivial. Most variable stars are more
difficult to be classified in the near-infrared bands, especially in
$K_s$, because their light curves contain less features than in the
optical (e.g., fundamental-mode RR~Lyrae show smaller amplitudes and
more sinusoidal light curves in the near-infrared than in the
optical), and there is a general absence of near-infrared template
light curves, compared to the situation in the optical. Within the VVV
collaboration we are trying to solve this shortcoming with the VVV
Templates Project \citep{ang14}, which will eventually lead to
automated classification of all VVV light curves \citep{cat13}. Since
we are still developing the algorithms for such an automated classification, for the present study we adopted an
''eyeball'' classification, providing a type only for those variables that had
well-measured light curves and could be securely classified based on
their periods and characteristic light curve
features. Fundamental-mode RR~Lyrae stars, Cepheids, eclipsing
binaries (mostly of Algol-type), and long-period variables (LPVs), are
found among the stars that we believe could be reliably classified in
this way.

\section{2MASS-GC02}
\label{sec_2ms02}
2MASS-GC02 is an inner Galactic GC, only recently discovered by
\citet{hur00} serendipitously during a spot check of 2MASS images for
quality assurance review. Its elevated extinction made it invisible in
earlier optical observations.  Since its discovery, there have been a
few photometric studies trying to determine its physical parameters
\citep{iva00,bor02,bor07}, with a range of values proposed for
extinction and distances, depending on various metallicity and
chemical enrichment scenarios. \citet{iva00} quoted values
$(m-M)_0=12.98$ and $E(J-K)=2.93$ for ${\rm [Fe/H]}=-0.5$, $(m-M)_0=13.70$ and
$E(J-K)=2.81$ for ${\rm [Fe/H]}=-1.0$, or $(m-M)_0=14.43$ and $E(J-K)=2.70$ for
${\rm [Fe/H]}=-2.0$, while \citet{bor07} found $(m-M)_0=13.60$ and
$E(J-K)=2.98$ for a disk-like enrichment scenario, or $(m-M)_0=13.48$ and
$E(J-K)=3.01$ for a bulge-like one.

To start our analysis of 2MASS-GC02, we built the cluster CMDs for the
various available filter combinations from our VVV photometry. In
Figure~\ref{fig_cmd2ms02}, we show the $K_s$ vs. $J-K_s$ CMD.  Stars
inside the half-light radius, plotted with bigger solid squares, allow us to
distinguish the red-giant branch (RGB) of the cluster, with a clump at
$K_s\approx14$ and $J-K_s\approx3.2$, but do not reach the
main-sequence (MS) turn-off point. In addition, 2MASS-GC02 is
poorly-populated, and it lies in a region of high field stellar
densities, so including in the CMD stars located farther away from the
cluster center, but still well within its tidal radius~-- which in
other cases, like halo GCs surrounded by fields with small stellar
densities, would help to increase the definition of the RGB branch~--,
here only makes the task of analyzing the cluster CMD more
difficult. The elevated number of bulge field giants present, lying in
the same CMD region where the cluster stars are located, adds
confusion to the study of 2MASS-GC02. The presence of significant
differential extinction in the field, clearly shown by the broadenings
in the RGB and red clump (RC) of the bulge stars present, complicates
even more the CMD analysis. To complete the description of the
observed CMD, we should mention that a significant number of field,
main-sequence disk stars can be observed as a clear branch of stars
bluer than $J-K_s<2$.  A deeper, more quantitative analysis of the
cluster CMD will probably not allow us to extract the physical
parameters of the cluster with better accuracy than previous
photometric studies down to similar magnitude depths
\citep{iva00,bor07}. Instead, we decided to take advantage of our
multi-epoch variability survey and use it to extract these physical
parameters from the information provided by the RR~Lyrae contained in
the cluster.

\subsection{Variables in 2MASS-GC02}
\label{sec_var2ms02in}
\begin{figure*}
\plotone{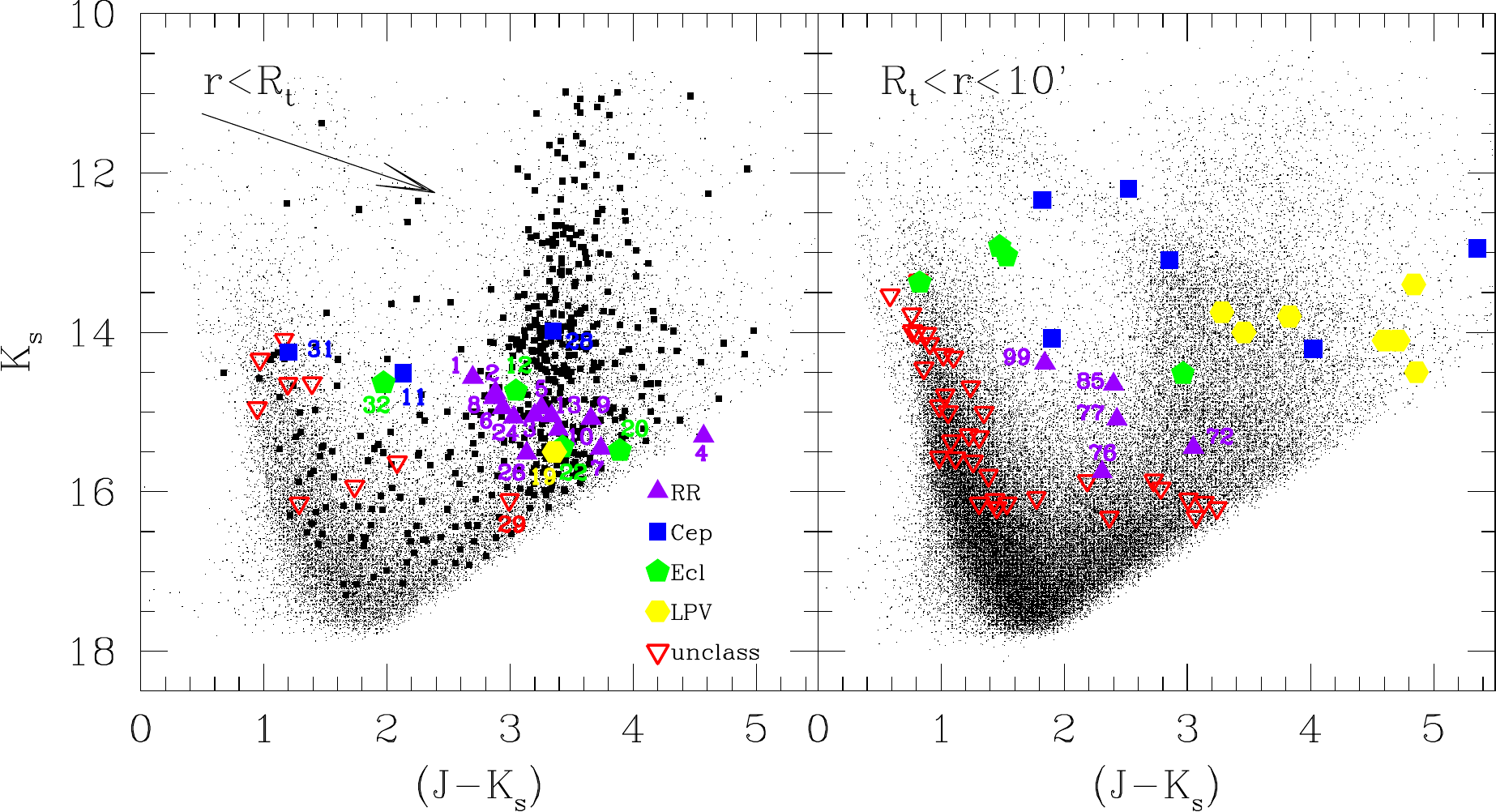}
\caption{$J-K_s$ vs. $K_s$ CMDs of 2MASS-GC02, out to its tidal radius
  $r_t=4.9\arcmin$ (left), and of its surrounding region (right). The arrow
  shows the reddening vector according to \citet{nis09}. In the left
  panel we have plotted with bigger solid squares the objects out to
  the cluster's half-light radius, $r_h=0.55\arcmin$. We have also
  overplotted the positions of RR~Lyrae as solid magenta triangles,
  Cepheids as solid blue squares, eclipsing binaries as solid green
  pentagons, LPVs as solid yellow hexagons, and unclassified variable
  candidates as red open triangles. See the electronic edition of the
  Journal for a color version of this figure*.}
\label{fig_cmd2ms02}
\end{figure*}

\begin{figure*}
\plotone{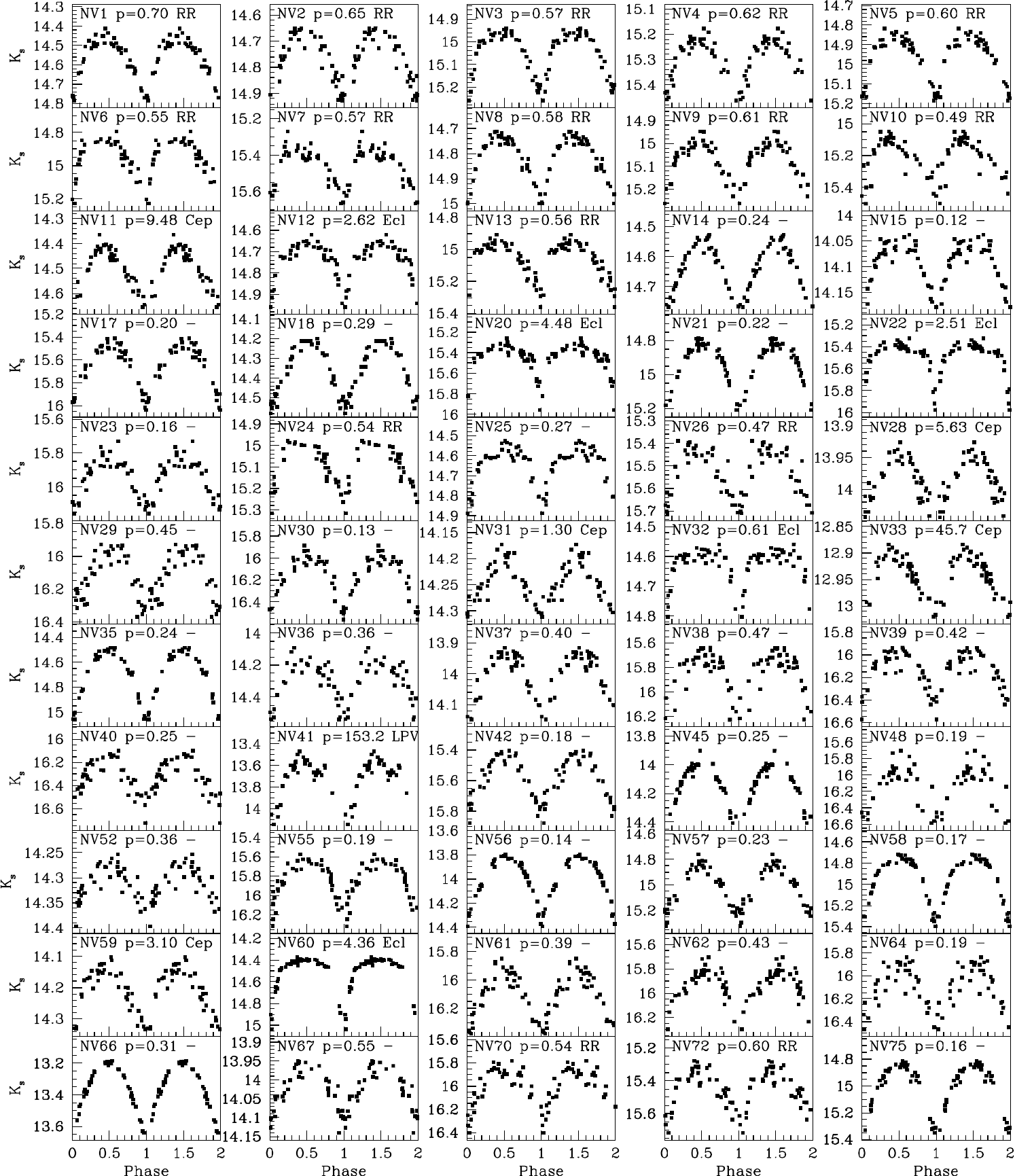}
\caption{Phase-folded light curves of the variable candidates in
  2MASS-GC02 and surroundings. The ID, the (rounded) period in days,
  and the variable type (where available), are indicated at the top of
  each panel. Light curves for the LPV candidates with periods longer
  than 500 days are not phase-folded. Supplemental data of this figure*
  is available in the online journal.}
\label{fig_lc2ms02}
\end{figure*}

There are observations in $K_s$ in 43 different epochs for 2MASS-GC02
during the period 2010-2013 of the VVV survey (see
Table~\ref{tab_obs}), although we only considered 42 in our following
analysis since tracking problems during the observations of one set
produced very elongated stars that were unusable for our variability
study. We more than triple the number of epochs in the only
previous variability study of this cluster, performed by
\citet{bor07}, with only 13 epochs. They found 5 RR~Lyrae variable
candidates, but our improved photometry and temporal coverage shows
that none of them is a true variable star\footnote{Although our NV2 and
  their V3 RR~Lyrae could be in principle cross-matched, they are
  already $3.1\arcsec$ away, i.e., separated by more than 9 pixels in the
  VVV images, and the reported periods and average magnitudes are
  quite different.}. Fortunately, after performing the variability
analysis described in Section~\ref{sec_varana}, we were able to find 32
new variable candidates inside the tidal radius. Among them, we
identified 13 RR~Lyrae variables, 3 Cepheids, 4 eclipsing binaries,
and 3 LPVs. Their positions, periods, amplitudes, magnitudes and
colors are shown in Table~\ref{tab_var2ms02} (NV1 to NV32), their light
curves are plotted in Figure~\ref{fig_lc2ms02}, their positions on the
sky can be observed in Figure~\ref{fig_chart2ms02}, and their
positions in the CMD can be observed in the left panel of
Figure~\ref{fig_cmd2ms02}.

All of the RR~Lyrae candidates found (NV1 to NV10, NV13, NV24 and
NV26) seem to be RRab's according to their periods and light
curves. From their position in the CMD and on the sky, it is
immediately clear that the RR~Lyrae candidates towards the north of
the cluster center (NV4, NV7, NV9 and NV13) have redder colors, and
therefore seem to be more severely affected by extinction, than the
RRab candidates in the south (NV1, NV2, NV5, NV6, NV8). This feature
points towards significant differential extinction across the face of
the cluster.  From their position in the different available CMDs, all
the RRab candidates seem to belong to the cluster except NV26~-- we
will further confirm this point in Section~\ref{sec_dis2ms02}. NV26 is
approximately half a magnitude dimmer than the other RRab variables
with similar colors, which implies that the increase in magnitude is
not due to higher reddening. This feature, clearly observable in
Figure~\ref{fig_ext2ms02}, along with being the one farthest away from
the cluster center and having a period significantly shorter than most
of the other RRab candidates, suggest it is a background field
RR~Lyrae.

The Cepheid candidates NV11, NV28, and NV31 show periods between
$\approx1$ and $\approx10$ days, and light curves with a sawtooth
shape, characteristic of these stars in $K_s$
\citep{ang14}. Unfortunately, none of them seem to have high chances
to belong to the cluster. NV11 and NV28 are too dim to be Type II
Cepheids belonging to 2MASS-GC02. According to the period-luminosity
relations by \citet{mat13}, their apparent distance moduli are
$\mu_{\rm NV11}=17.80$ and $\mu_{\rm NV28}=16.76$, which differ
significantly from those of the RR~Lyrae shown in
Figure~\ref{fig_ext2ms02}~-- the method to obtain those is explained
in Section~\ref{sec_dis2ms02}. Their position in the CMD suggest they
are background bulge field Cepheids. On the other hand, NV31 has
$\mu_{\rm NV31}=15.57$, according to the period-luminosity relations by
\citet{mat13} for Type II Cepheids. This apparent distance modulus is
in the range shown by the RR~Lyrae in the GC (see
Figure~\ref{fig_ext2ms02}), but as shown in the CMD in
Figure~\ref{fig_cmd2ms02}, its color is too blue for NV31 to belong to
the cluster or bulge.

The eclipsing binary candidates NV12, NV20, NV22, and NV32 have light
curve characteristics of Algol-type binaries (EA). NV32 is too blue to
be a cluster member based on its position in the CMD, but the
remaining EA candidates cannot be disregarded as cluster members from
the CMD examination since they lie in regions where there are cluster
stars.

At the present time, there is not much we can say about the 3 LPVs
discovered, NV16, NV19 and NV27. Their long periods have not allowed us
yet to sample their light curves with enough phase coverage for an
accurate period determination. Also their red colors seem to be the
reason NV16 has been detected only in $K_s$, NV27 in $H$ and $K_s$, and
NV19 in $J$, $H$ and $K_s$.

Almost all of the remaining 9 unclassified variables seem to be disk
field stars based on their position in the CMD. The only one that
could be a cluster member is NV29.

\subsection{Variables surrounding 2MASS-GC02}
\label{sec_var2ms02out}
\begin{figure*}
\plotone{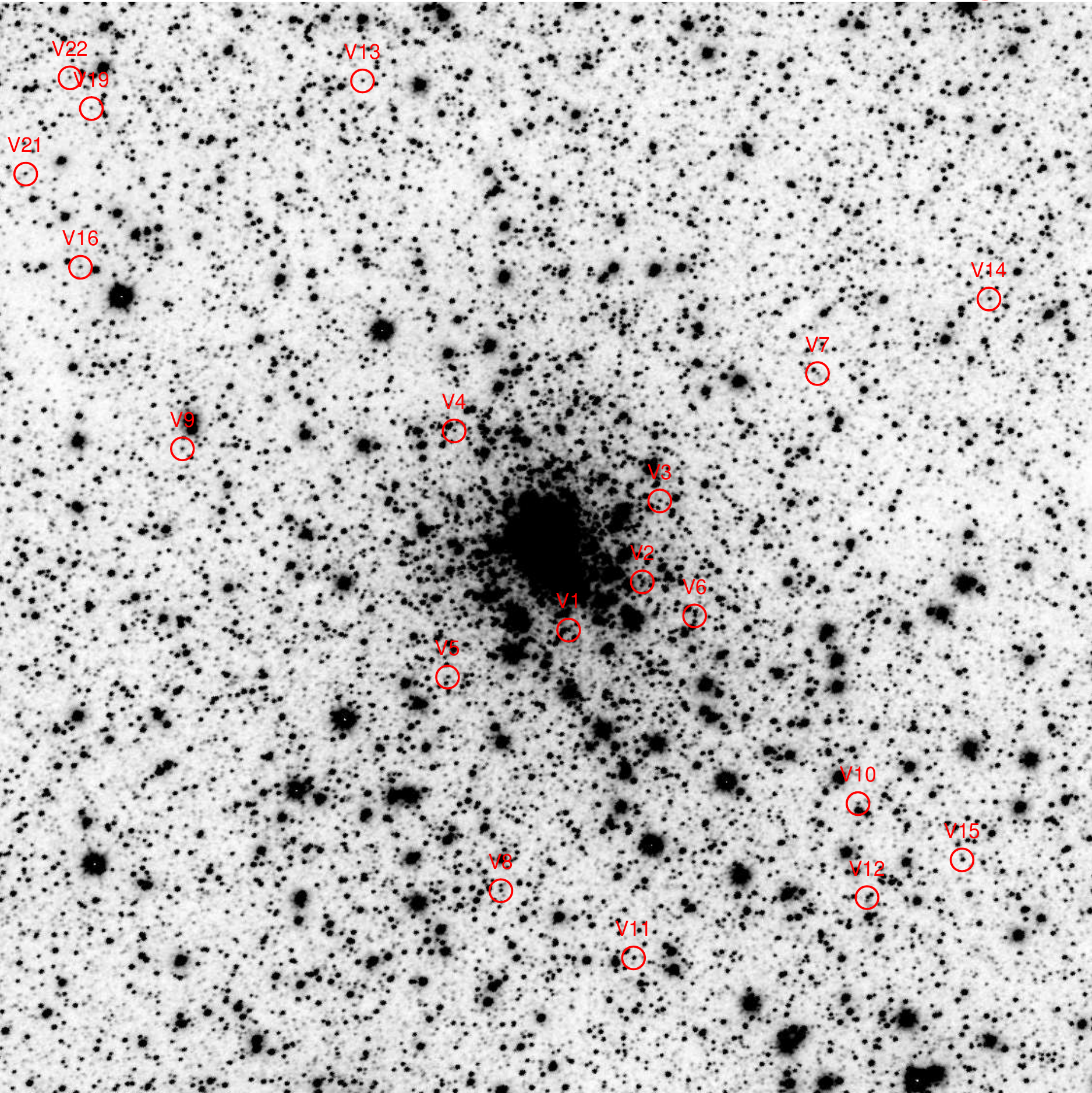}
\caption{$K_s$-band finding charts for the variables found in
  2MASS-GC02 and surroundings. North is up, east is right. The field
  of view is $6\arcmin\times6\arcmin$, and includes most of the RR~Lyrae in the
  cluster. A finding chart with all the variables found (NV1 to NV102)
  is provided additionally only in the electronic edition.}
\label{fig_chart2ms02}
\end{figure*}

As previously mentioned in Section~\ref{sec_obs}, we performed a
search for variables out to a radius of $10\arcmin$ from the cluster
center. This allowed us to look for variables in the field surrounding
the GC, in an area approximately 3 times the size of the cluster. We
found a total of 70 new variable candidates in this surrounding
region. Their positions, periods, amplitudes, magnitudes and colors
are shown in Table~\ref{tab_var2ms02} (NV33 to NV102), their light
curves are plotted in Figure~\ref{fig_lc2ms02}, their positions on the
sky can be observed in Figure~\ref{fig_chart2ms02}, and their
positions in the CMD can be observed in the right panel of
Figure~\ref{fig_cmd2ms02}.

We found 6 extra-tidal RR~Lyrae (NV70, NV72, NV76, NV77, NV85, and
NV99) in the region studied outside the tidal radius. From the shape
of their light curve and their periods they are RRab's. From their
position in the CMD all of them are clearly in the background of the
GC, except NV85 and NV99~--we will further confirm this point in
Section~\ref{sec_dis2ms02}--. The high angular separations from
the center of the cluster (see Table~\ref{tab_var2ms02}) and the
positions in the sky (see Figure~\ref{fig_chart2ms02}) of NV85 and
NV99 suggest that, instead of being variables in a hypothetical GC
tidal tail, they have a higher chance of being field stars located at
a distance similar to that of the cluster. Most of the found field
RR~Lyrae being in the background of the cluster, points towards
2MASS-GC02 being in front --or in the near-side-- of the Galactic
bulge.

In addition to these 6 RR~Lyrae, we also identified 6 Cepheids, 4
eclipsing binaries, and 19 LPV candidates.  From a purely
statistical point of view, being the surrounding area 3 times bigger than the region inside the tidal radius, we should expect 2 field RR~Lyrae, 2 field Cepheids, 1 field eclipsing binaries, and 6 field LPV candidates inside the cluster radius. These numbers of variables are within the range of these objects found in Section~\ref{sec_var2ms02in}, given that we are dealing with small number statistics, except for the case of RR~Lyrae, where we find many more (however, it agrees with the number of field RR~Lyrae found in Section~\ref{sec_var2ms02in}). Therefore, this strengthens our assumption that most of the inner RR~Lyrae candidates found in Section~\ref{sec_var2ms02in} are cluster members, and most, if not all, of the inner Cepheids, eclipsing binaries, and LPV candidates are field stars. 

\subsection{Distance, reddening and Oosterhoff type}
\label{sec_dis2ms02}
The reddening values and distances of every individual RRab star can
be accurately calculated using their tight
period-luminosity-metallicity relations in the near-infrared
\citep{lon86,lon90,bon01,cas04,cat04}. Following the steps detailed in
\citet{dek13}, we adapted the \citet{cat04} period-luminosity
relations to be used with the VIRCAM/VISTA filter system:
\begin{equation}
M_{K_s} = -0.6365 - 2.347 \log(P) + 0.1747 \log(Z)
\end{equation}
\begin{equation}
M_H  = -0.5539 - 2.302 \log(P) + 0.1781 \log(Z)
\end{equation}
\begin{equation}
M_J  = -0.2361 - 1.830 \log(P) + 0.1886 \log(Z)
\end{equation}
\begin{equation}
M_Y  = +0.0090 - 1.467 \log(P) + 0.1966 \log(Z)
\end{equation}
\begin{equation}
M_Z  = +0.1570 - 1.247 \log(P) + 0.2014 \log(Z)
\end{equation}
where $P$ is the period of the RRab and $Z$ its metallicity.

The only spectroscopic determination of the iron content for the stars
in 2MASS-GC02 we have found in the literature is by
\citet{bor07}. They found a value of ${\rm [Fe/H]}=-1.08$, using
low-resolution infrared spectroscopy of secondary lines in a set of 12
cluster stars, selected by their photometry and radial
velocities\footnote{Pe\~naloza et al. (in preparation) also
  report a value of ${\rm [Fe/H]}=-1.08$ from high-resolution spectra of a
  few cluster giants.}. Although photometric estimations are a little
higher, with values between ${\rm [Fe/H]}=-0.66$ \citep{bor02} to
${\rm [Fe/H]}=-0.98$ \citep{bor07}, we decided to stick to the spectroscopic
value, which is also the one provided by the \citet{har96} catalog
(see Table~\ref{tab_gc}). Using a canonical helium fraction $Y=0.245$,
and an alpha-element enhancement $[{\alpha}/Fe]=0.3$, common among
Galactic GCs \citep{pri05}, we translate the iron content
${\rm [Fe/H]}=-1.08$ to the metallicity $Z=0.0025$ that we will assume for
all the RRab stars in the cluster.

\begin{figure*}
\epsscale{1}
\plotone{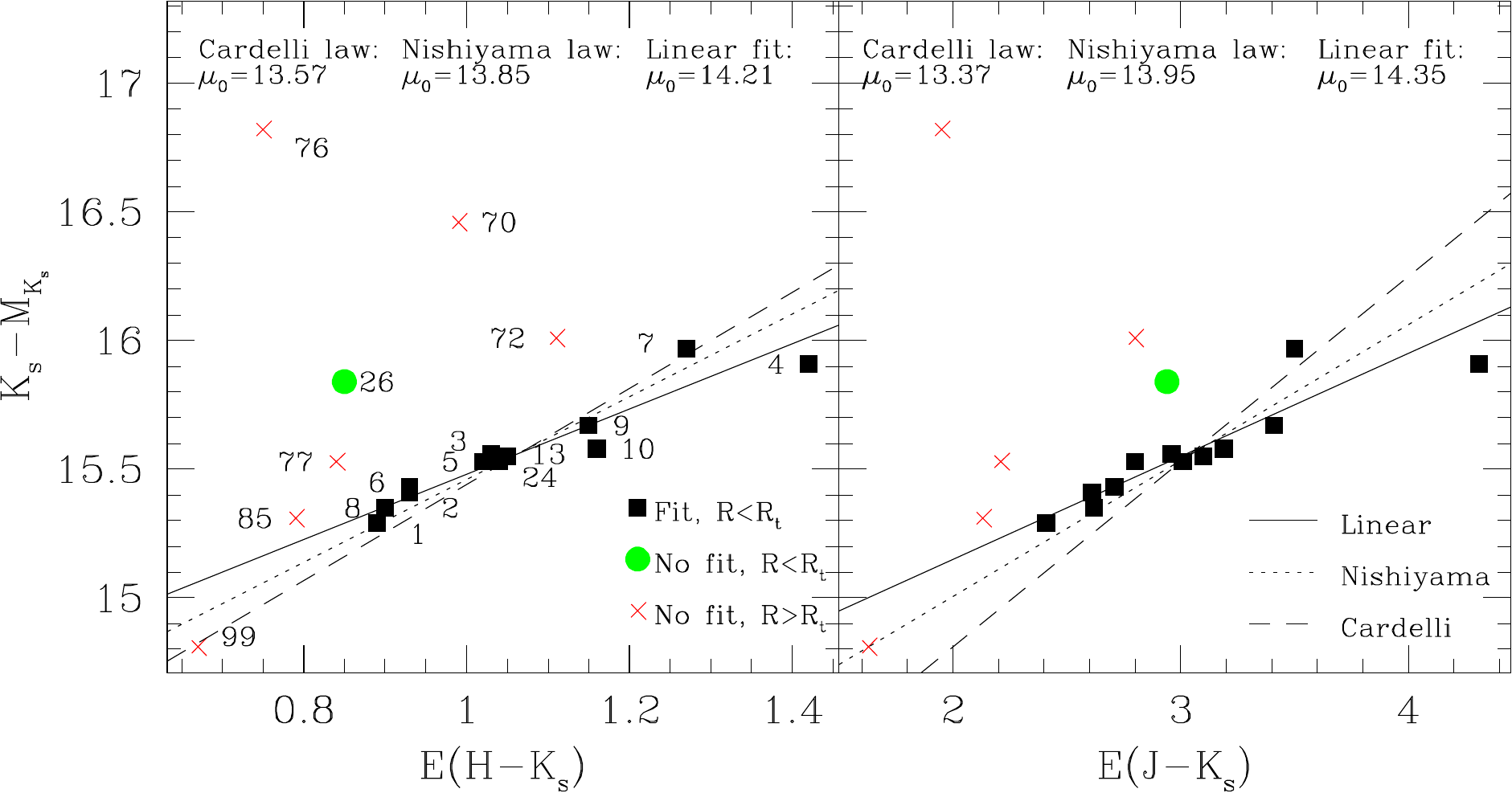}
\caption{Apparent distance modulus vs. color excess diagrams for the
  RR~Lyrae in 2MASS-GC02 and surroundings. A linear fit (solid line)
  to the cluster RRab variables will give the extinction law (slope of
  the fit) and the absolute distance modulus (zero-point of the fit)
  of the cluster. Black squares are the cluster RRab candidates used
  in the fit, green circles are RRab candidates inside the tidal
  radius not used in the fit because of their high residuals, and red
  crosses are RR~Lyrae in the surrounding field. As a comparison, we
  have also plotted the fit using the \citet{car89} extinction law
  (dashed line), and the \citet{nis09} extinction law (dotted line).}
\label{fig_ext2ms02}
\end{figure*}

It is straightforward to calculate the apparent distance modulus of
the RRab candidates in $K_s$ (see Figure~\ref{fig_ext2ms02}), using
the calculated period and average apparent magnitudes in
Table~\ref{tab_var2ms02}. But to calculate the absolute distance
modulus we need to carry out some further calculations and adopt some
further assumptions in order to obtain the extinction first. Ideally
to calculate the extinction, we should also have the light curve in
some other near-infrared filter in addition to $K_s$, but VVV only
provides single-epoch observations in $Z$, $Y$, $J$, and $H$. What is
worse, the $Z$ and $Y$ magnitudes of the RR~Lyrae candidates in
2MASS-GC02 are in most cases below the detection limit (see
Table~\ref{tab_var2ms02}). For our analysis, we assume that $J-K_s$
(or $H-K_s$) at the phase position when the $J$ (or $H$) image was
taken is equal to the average $J-K_s$ (or $H-K_s$) color of the
RR~Lyrae candidate. We obtain the $K_s$ value at the desired phase
from our Fourier fits to the light curves for the RRab candidates (see
Section~\ref{sec_varana}), calculate this way the apparent colors of
these RR~Lyrae candidates, shown in Table~\ref{tab_var2ms02}, and
using equations 3 to 5 obtain their color excesses $E(J-K_s)$ and
$E(H-K_s)$, shown in Figure~\ref{fig_ext2ms02}.

Now, in order to obtain the absolute distance modulus to 2MASS-GC02 we
should transform those color excesses to extinctions using the
total-to-selective extinction ratios. But these ratios have been shown
to differ from the values obtained from the standard \citet{car89}
extinction law, when observing towards the Galactic center and bulge
\citep[e.g.][]{nis06,nis09}. Therefore, instead of adopting a ratio
from the literature, the highly differential reddening inside the
cluster's area enables us to simultaneously obtain the absolute
distance modulus and the total-to-selective extinction ratio (and
hence the extinction) by doing a linear fit between the two parameters
we already had, the apparent distance modulus $K_s-M_{K_s}$ and the
color excess ($E(J-K_s)$ and $E(H-K_s)$). Since there are errors
associated with both variables, the fit is performed using the
ordinary least square bisector method, which has been proven to
outperform other approaches in such cases \citep{iso90,kun08}.  The
fits are shown in Figure~\ref{fig_ext2ms02}. As we can expect, most
RR~Lyrae outside the tidal radius are obvious outliers in the fit, and
can be clearly located in the background of the cluster (see
Section~\ref{sec_var2ms02out}). From the RR~Lyrae inside the tidal
radius, only NV26 has been also considered to be an outlier, due to
its considerable residuals, and was discarded from the fitted sample
(see discussion and explanations for these cases in
Section~\ref{sec_var2ms02in}). From the final fit, we obtain the
ratios $R_{K_s,J-K_s}\equiv {\cal A}_{K_s}/(E(J-K_s)$ and
$R_{K_s,H-K_s}\equiv {\cal A}_{K_s}/(E(H-K_s)$ shown in
Table~\ref{tab_ext}.  These ratios are a little lower than those
obtained by \citet{nis09}, and are certainly smaller than suggested by
\citet{car89} or \citet{rie85}. The absolute distance modulus,
calculated as the average of the zero-points of the fits, is
$\mu_{0}=14.26\pm0.14\pm0.3$, where the first ${\sigma}$ is obtained from the
statistical errors in the fit, and the second one comes from omitting
some of the fitted stars, as we explained in the following paragraph. This
value for the absolute distance modulus yields the cluster's distances
to the Sun and to the Galactic center shown in Table~\ref{tab_dis},
putting the object farther away from us and more than 1 kpc closer to the
Galactic center than shown in the \citet{har96} catalog. For the sake
of comparison, we chose as a good reference value for the extinction
towards 2MASS-GC02 $E(J-K_s)=3.1\pm0.5$, the mean of the cluster RRab
candidates, but we would like to remark that extinction towards this
cluster is highly variable, with almost 50\% changes over small regions, and
with significant deviations from the standard extinction law.

The assumption of ${\cal A}_{\lambda}/{\cal A}_{K_s}=1$ that we have done by using only one epoch in J and H in order to obtain the colors of the RR~Lyrae, does not add significantly to the error budget of the different
calculated magnitudes. Using the relations from Appendix B1.3 in
\citet{fea08}, we observed that the ratios between the amplitudes of
the RR~Lyrae in the near-infrared change according to the amplitudes
of the RR~Lyrae, and for the amplitudes of the detected RR~Lyrae (see
Table~\ref{tab_var2ms02}), the values\footnote{A similar value for
  ${\cal A}_J/{\cal A}_{K_s}$ is obtained by Navarrete et al. (in
  preparation) when studying the RR~Lyrae light curves in the
  near-infrared in OmegaCen.} should be in the ranges
$0.9<{\cal A}_H/{\cal A}_K<1.1$ and $0.8<{\cal A}_J/{\cal A}_K<1.9$,
and the errors in the RR~Lyrae colors due to our assumption could be,
at most, of one tenth of a magnitude in $J-K_s$, and one hundredth of
a magnitude in $H-K_s$. Note however that these would be the extreme
cases when the measurements were taken at a phase corresponding to a
maximum or a minimum; generally the error should be smaller, and their
effect on the elevated color excesses that we obtained (see
Figure~\ref{fig_ext2ms02}) almost negligible.  What is more, any effects from our
assumption of ${\cal A}_{\lambda}/{\cal A}_{K_s}=1$ would cancel out
to some extent in the fit we have done to obtain the absolute distance
modulus and the reddening law, since the measurements of $J$ and $H$
were done at random phase times, which means that errors in color
excesses should have similar chances of being positive and negative,
and the only observable effect would be a small increase in the
dispersion of the points around the fit. More important for the error
budget of the extinction ratios $R$ and the absolute distance modulus
$\mu_{0}$ would be if one of the fitted RR~Lyrae with a higher
extinction (NV4 or NV7) resulted not to be a member of the
cluster. Since they present the highest deviations from the fit, and
they are also important to extend the extinction baseline, we decided
to show the effects of omitting one of them as the
variations shown by the second term in the error of the extinction
ratios $R$ and the absolute distance modulus $\mu_{0}$.

\begin{figure*}
\plotone{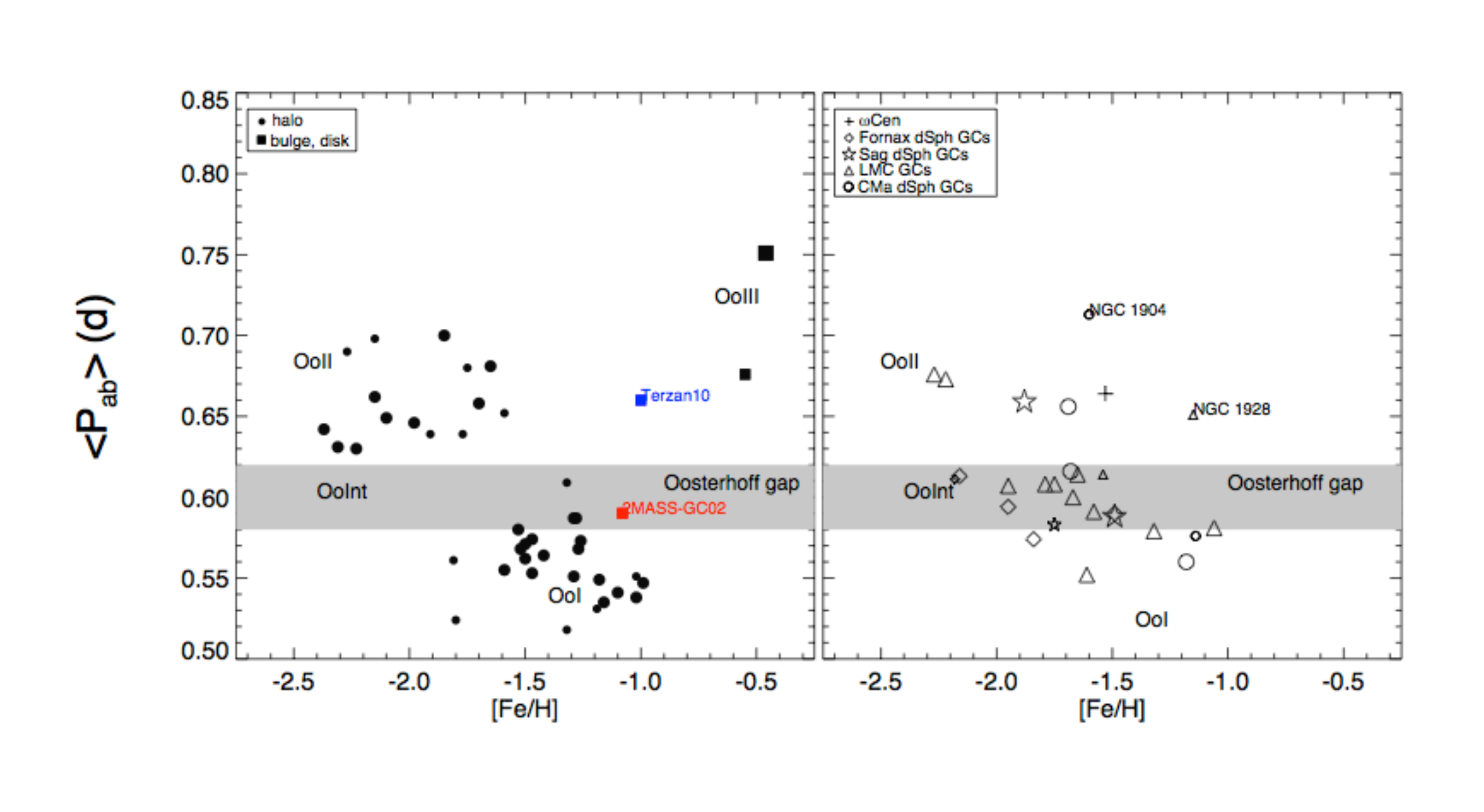}
\caption{Distribution of Galactic GCs (left) and stellar populations
  associated with neighboring dwarf galaxies (right) in the average
  ab-type RR~Lyrae period $\langle P_{\rm ab} \rangle$
  vs. ${\rm[Fe/H]}$ plane, updated from Figure 5 in \citep{cat09a}. Smaller
  symbols indicate that an object contains between 5 and 10 RRab's,
  while bigger symbols indicate more than 10 RRab's. Locations of
  2MASS-GC02 and Terzan~10 have been plotted with different colors, to
  highlight the unusual positions of both studied GCs in the general
  Galactic GCs distribution.}
\label{fig_oos}
\end{figure*}

The average period of the cluster RRab candidates used in the fit
(NV1-NV10, NV13 and NV24) is $\langle P_{ab} \rangle
=0.59\pm0.06$ days, which puts 2MASS-GC02 in the Oosterhoff gap, as we can
see in Figure~\ref{fig_oos}. Based on the absence of Galactic GCs in
this gap, and their high incidence in nearby extragalactic systems
(see Figure~\ref{fig_oos}), this feature could suggest an
extragalactic origin for 2MASS-GC02. However, we should note that we
are dealing with small number statistics here, since only 12 RRab's
are considered.

\section{Terzan~10}
\label{sec_t10}
Since its discovery by \citet{ter71}, there have been only a few
photometric studies published on Terzan~10~-- e.g., \citet{liu94} in
the near-infrared, \citet{ort97} in the optical; and no spectroscopic
studies of its individual stars, although \citet{bic98} studied the
integrated spectra of this GC. Estimations of the reddening range from
$E(B-V)=1.71$ \citep{web85} or $E(B-V)=1.90$ \citep{bic98} to
$E(B-V)=2.40$ \citep{ort97} or $E(B-V)=2.60$ \citep{liu94}. Its
distance modulus estimations also show a significant range, varying
from $(m-M)_0=13.40$ \citep{ort97} to $(m-M)_0=14.5$ \citep{liu94}.

\begin{figure*}
\plotone{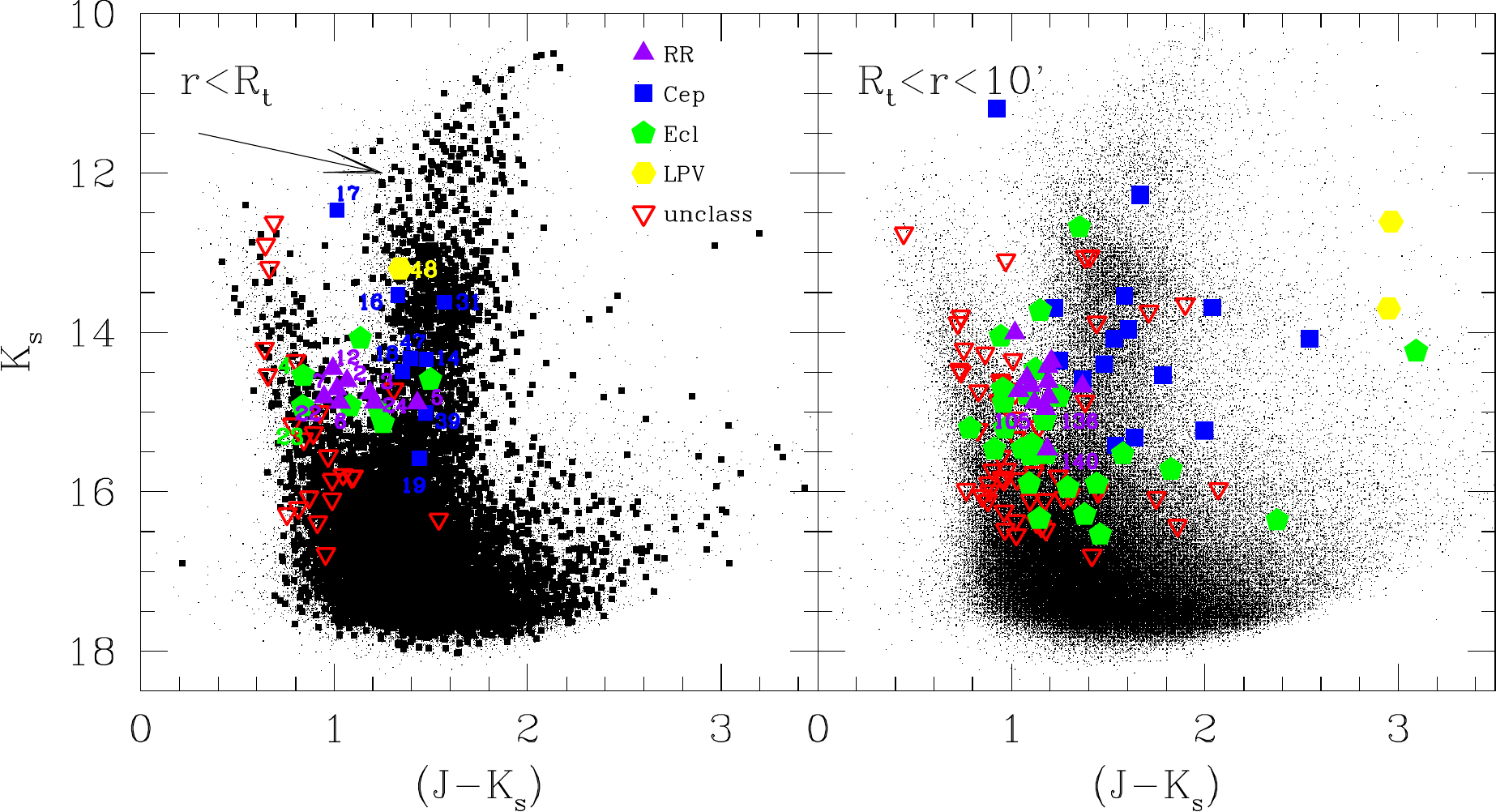}
\caption{As in Figure~\ref{fig_cmd2ms02}, but for Terzan~10 and its
  surroundings. In the left panel we have plotted with bigger solid
  squares the objects out to the cluster's half-light radius, $r_h=1.55\arcmin$.}
\label{fig_cmdt10}
\end{figure*}

In Figure~\ref{fig_cmdt10}, we show the $K_s$ vs. $J-K_s$ CMD of the
cluster that we built with our VVV PSF photometry. It is immediately
clear that, although extinction is lower than for 2MASS-GC02 (see
Figure~\ref{fig_cmd2ms02}), it is difficult to disentangle cluster
stars from their field counterparts. Even at distances smaller than
the half-light radius, plotted with bigger solid squares in the left
panel of Figure~\ref{fig_cmdt10}, there is a significant presence of
disk and bulge field populations in addition to the GC. The RGB of the
cluster, with a clump at $K_s\approx13.5$ and $J-K_s\approx1.5$, is
mixed with bulge field giants. The brightest disk field stars are
bluer than the upper RGB stars ($J-K_s<1.2$) and can be easily
separated from them, but the dimmest disk field stars contribute to
the confusion caused by the bulge field stars, since both are located
in the same region as the subgiant branch (SGB) and upper MS of the
cluster. Therefore, even though our photometry reaches the turn-off
point of the MS, we cannot properly define it due to the contamination
by the field populations. Again, instead of trying to perform a deeper
CMD analysis and statistical field star decontamination, we decided to
take advantage of our variability survey and use it to extract
Terzan~10 physical parameters from the information provided by the
RR~Lyrae variables we identified in the cluster.

\subsection{Variables in Terzan~10}
\label{sec_vart10in}
\begin{figure*}
\plotone{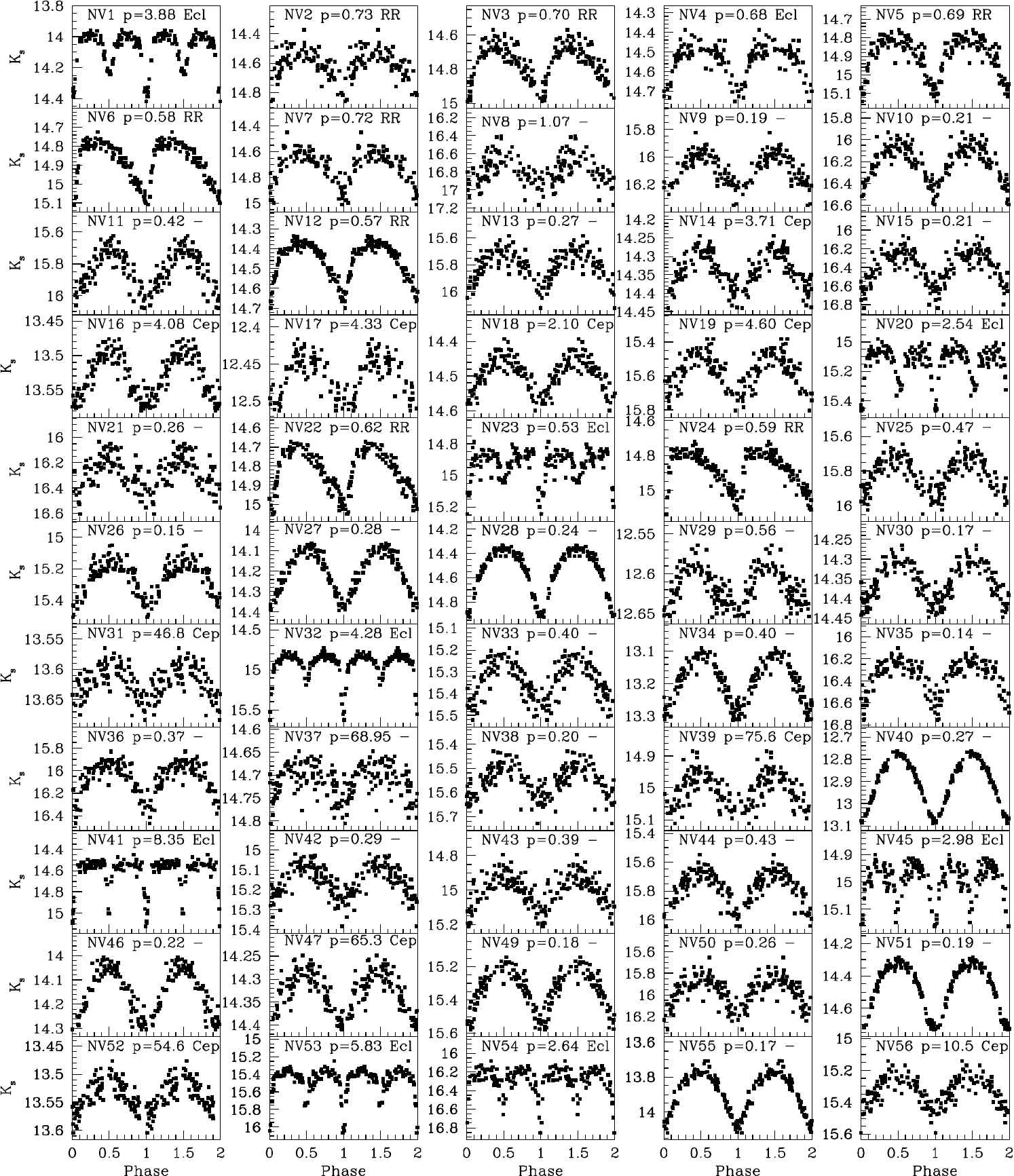}
\caption{As in Figure~\ref{fig_lc2ms02}, but for Terzan~10 and its
  surroundings. Supplemental data of this figure is available in the
  online journal.}
\label{fig_lct10}
\end{figure*}

\begin{figure*}
\plotone{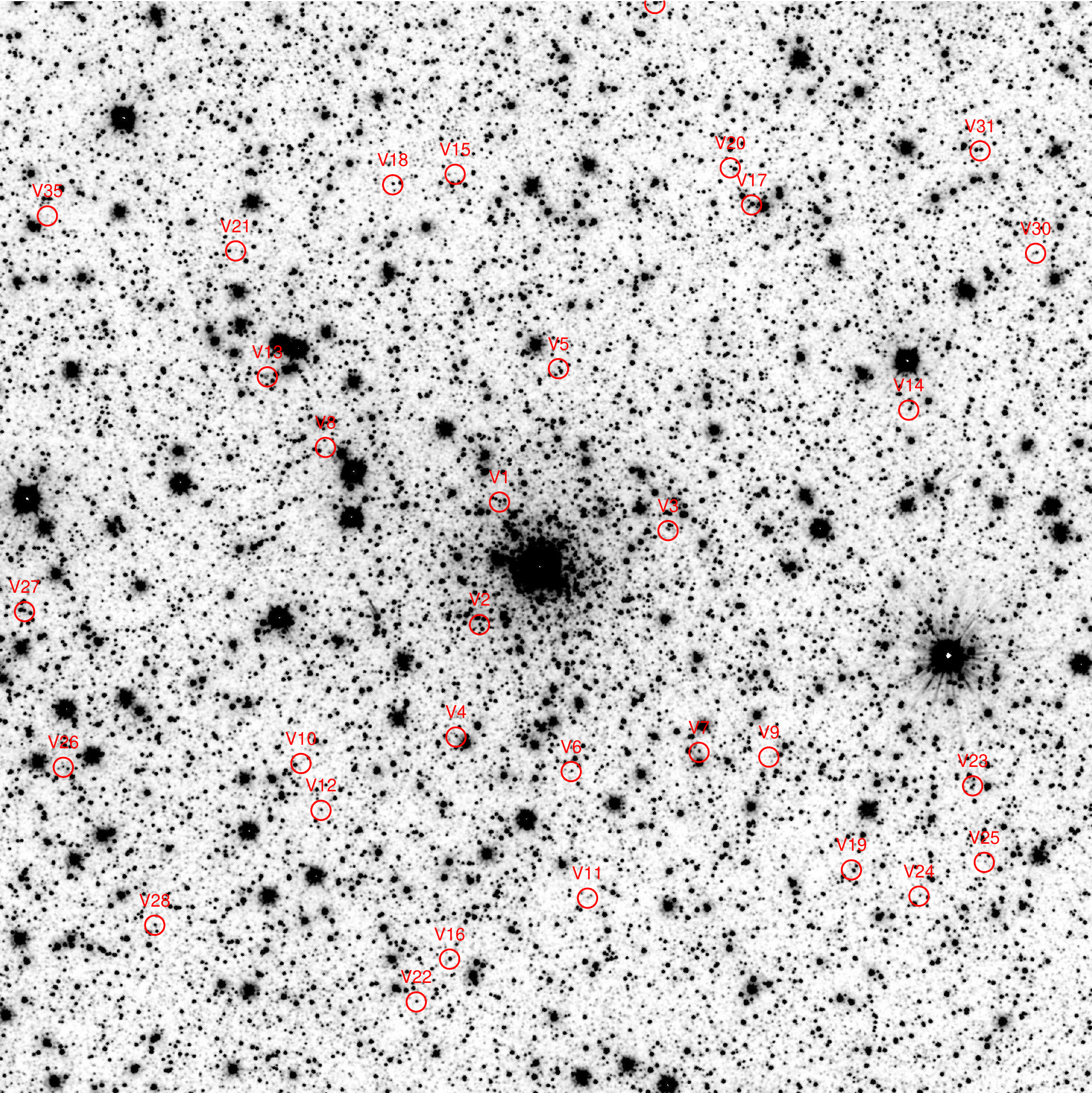}
\caption{As in Figure~\ref{fig_chart2ms02}, but for Terzan~10.}
\label{fig_chartt10}
\end{figure*}

Observations in $K_s$ were taken in 101 different epochs for Terzan~10
during the period 2010-2013 of the VVV survey (see
Table~\ref{tab_obs}), making this region one of the most sampled in
the VVV so far. No previous search for variable stars in this cluster
has been ever done\footnote{During the revision process of our article, \citet{sos14} published their results on the RR~Lyrae in the OGLE Galactic Bulge fields. Since Terzan~10 falls on one of the fields covered by OGLE-IV, we matched their results with ours, and found that most of the RRab's in the field of Terzan~10 and in its surroundings appear in both works. There are only 3 RRab candidates inside the cluster radius of Terzan10 (OGLE\_BLG\_RRLYR-33521, OGLE\_BLG\_RRLYR-33518, and OGLE\_BLG\_RRLYR-33525), and 1 RRab candidate in the surrounding region (OGLE\_BLG\_RRLYR-33508) that \citet{sos14} find, and we do not. The reason for this is that these stars lie very close to stars of similar or higher magnitude, and the image subtraction technique they use is better suited than our PSF photometry technique in this case. On the other hand, there are 1 RRab candidate inside the tidal radius of Terzan~10 (NV5) and 1 in the surrounding region (NV140) that we identify and they do not. The reason for this is their dimmer magnitudes in $I$, close to the magnitude limit where the completeness of the OGLE detection is small. We should also note that we are not able to recover any of the 7 RRc candidates that \citet{sos14} find in Terzan~10 and its surroundings, due to the smaller amplitudes of these variable stars in the near-infrared wavelengths.}, so this high amount of available epochs puts us in
a superb position to find for the first time highly significant
variable candidates in this GC. Our analysis resulted in 48 variable
candidates inside $r_t=5.06\arcmin$, the tidal radius extracted from the
\citet{har96} catalog. Among the 48 candidates, we identified 8
RR~Lyrae variables, 8 Cepheids, 7 eclipsing binaries, and 1 LPV. Their
positions, periods, amplitudes, magnitudes and colors are shown in
Table~\ref{tab_vart10} (NV1 to NV48), their light curves are plotted in
Figure~\ref{fig_lct10}, their positions in the sky can be observed in
Figure~\ref{fig_chartt10}, and their positions in the CMD can be
observed in the left panel of Figure~\ref{fig_cmdt10}.

All of the RR~Lyrae candidates found (NV2, NV3, NV5, NV6, NV7, NV12, NV22 and NV24)
seem to be RRab's, according to their periods and light curves. From
their position in the CMD and in the sky, the RR~Lyrae are located along
the reddening vector, suggesting the presence of differential
extinction in this GC, with the candidates towards the north of the
cluster center (NV3 and NV5) with redder colors, and therefore more
heavily affected by extinction, than the RRab candidates in the south
(NV2, NV6, NV7 and NV12).  On the CMD, we can observe that all
RR~Lyrae show a behavior consistent with their belonging to the
cluster and following the same, but again non-standard,
reddening law, except NV12~-- we will further confirm this
point in Section~\ref{sec_dist10}. NV12 is a little brighter than the other RR~Lyrae
with similar colors, which suggests that it is a foreground
RR~Lyrae star.

None of the Cepheid candidates found (NV14, NV16, NV17, NV18, NV19,
NV31, NV39 and NV47) could be a Cepheid that belongs to the cluster
according to the \citet{mat13} period-luminosity relations. While most
of them are too dim to even belong to the Galactic bulge, NV16 and NV17 have
apparent distance moduli consistent with being bulge Type II Cepheids
(${\mu}_{\rm NV16}=15.98$ and ${\mu}_{\rm NV17}=14.98$), along with
positions in the CMD accordant with it. But they do not seem to belong
to Terzan~10 either. A quick visual comparison with the RR~Lyrae
apparent distance moduli and extinctions shown in
Figure~\ref{fig_extt10}~-- the method to obtain them is explained in
Section~\ref{sec_dist10}~-- strongly points towards NV16 being a Type
II Cepheid in the background of the cluster, and NV17 being in its
foreground.

All the eclipsing binary candidates (NV1, NV4, NV20, NV23, NV32, NV41, and NV45)
lie in a region of the CMD compatible with being cluster members,
except NV4 and NV23, which are located among the disk field stars. Based on
their light curves most of them appear to be EA stars.

Only 1 LPV was found inside the tidal radius of the cluster, NV48. Its
long period precludes us from having a well-sampled light curve.

The vast majority of the remaining 23 unclassified variables, seems to
be comprised of disk field variables, based on their positions in the CMD (see
Figure~\ref{fig_cmdt10}).

\subsection{Variables surrounding Terzan10}
\label{sec_vart10out}
For Terzan~10, we also performed a search for variables in the
immediate surrounding of the cluster, out to a radius of $10\arcmin$ from
its center. This encircles an area approximately 3 times the size of
the cluster out to its tidal radius. In this region we found 112 more
variable candidates. Their positions, periods, amplitudes, magnitudes
and colors are shown in Table~\ref{tab_vart10} (NV49 to NV160), their
light curves are plotted in Figure~\ref{fig_lct10}, and their
positions in the CMD can be observed in the right panel of
Figure~\ref{fig_cmdt10}.

We found 12 RR~Lyrae candidates in the surveyed region, significantly
more than the RR~Lyrae found in the surrounding of 2MASS-GC02. But
extinction in that region was higher, as we can guess just by looking
at the redder colors of the stars there, and the angular separation
from the Galactic center was also higher, meaning that less of the
projected bulge was sampled. In general, RR~Lyrae found in the
surroundings of Terzan~10 are a little brighter than the ones found
inside its tidal radius, and their colors are similar, which suggests
similar reddenings. These two facts imply that they are field RR~Lyrae
in the foreground of the cluster, suggesting that the cluster is on
the far side of the bulge, farther away than the Galactic center. We
will confirm this result and further discuss it in
Section~\ref{sec_dist10}. From the CMD and from the further analysis
in Section~\ref{sec_dist10}, only one of the RR~Lyrae, NV140, 
is behind the cluster, and other two, NV105 and NV136, lie at a distance
consistent with being cluster members, although their high angular
distances from the cluster center make more probable that they are field stars at the cluster's distance. From a statistical point of
view, the expected number of field RR~Lyrae inside the cluster radius
should be $\approx4$, since the surrounding region is 3 times the area
occupied by the cluster. Although we only found 1 (see
Section~\ref{sec_vart10in}), we should note that we are dealing with
small number statistics here.

We also identified 15 Cepheids, 27 eclipsing binaries, and 2 LPVs. Therefore, we should expect in the area inside the cluster tidal radius close to 5 field Cepheids, 9 field eclipsing binaries and 1 field LPV, given the bigger size of the surrounding region (3:1). The
amounts of these kinds of variables 
found in Section~\ref{sec_vart10in} (8 Cepheids, 7 eclipsing binaries and 1 LPV) are of the same order, suggesting that most, if not all, of the variables of these
types found in Section~\ref{sec_vart10in} belong to the field
population, and not to the cluster.

The number of LPVs seems odd when compared with 2MASS-GC02. There we
found 19 LPVs, while in the surrounding of Terzan~10 we found only
2.

\subsection{Distance, reddening and Oosterhoff type}
\label{sec_dist10}
To calculate the distance and the extinction to Terzan~10 we follow
the same steps as in Section~\ref{sec_dis2ms02}. We use equations 1 to
5 to obtain the absolute magnitudes of our RR~Lyrae candidates in the
different available filters. Unfortunately, we found no spectroscopic
metallicity measurements of individual stars in Terzan~10, and we have
to rely on only photometric and integrated spectroscopic
values. Metallicity estimations vary between ${\rm [Fe/H]}=-0.7$ in
the photometric studies by \citet{liu94} to ${\rm [Fe/H]}=-1.2$
($[Z/Z_{\odot}]=-1.35$) in the integrated spectroscopic studied by
\citet{bic98}. We adopt ${\rm [Fe/H]}=-1.00$, from the \citet{har96}
catalog, and assuming the same $Y$ and $\alpha$-enhancement as in
Section~\ref{sec_dis2ms02}, we obtain a value of $Z=0.003$ for the
cluster's metallicity, that we used as input for equations 1 to 5,
along with our measured periods. The mean RR~Lyrae apparent $K_s$
magnitude is obtained as mentioned in Section~\ref{sec_varana}, and
the apparent colors from the $K_s$ measurements obtained by
interpolating in our Fourier fit to the $K_s$ light curve at the same
phase as the single-epoch images in $Z$, $Y$, $J$, and $H$ were taken.

\begin{figure*}
\epsscale{1}
\plotone{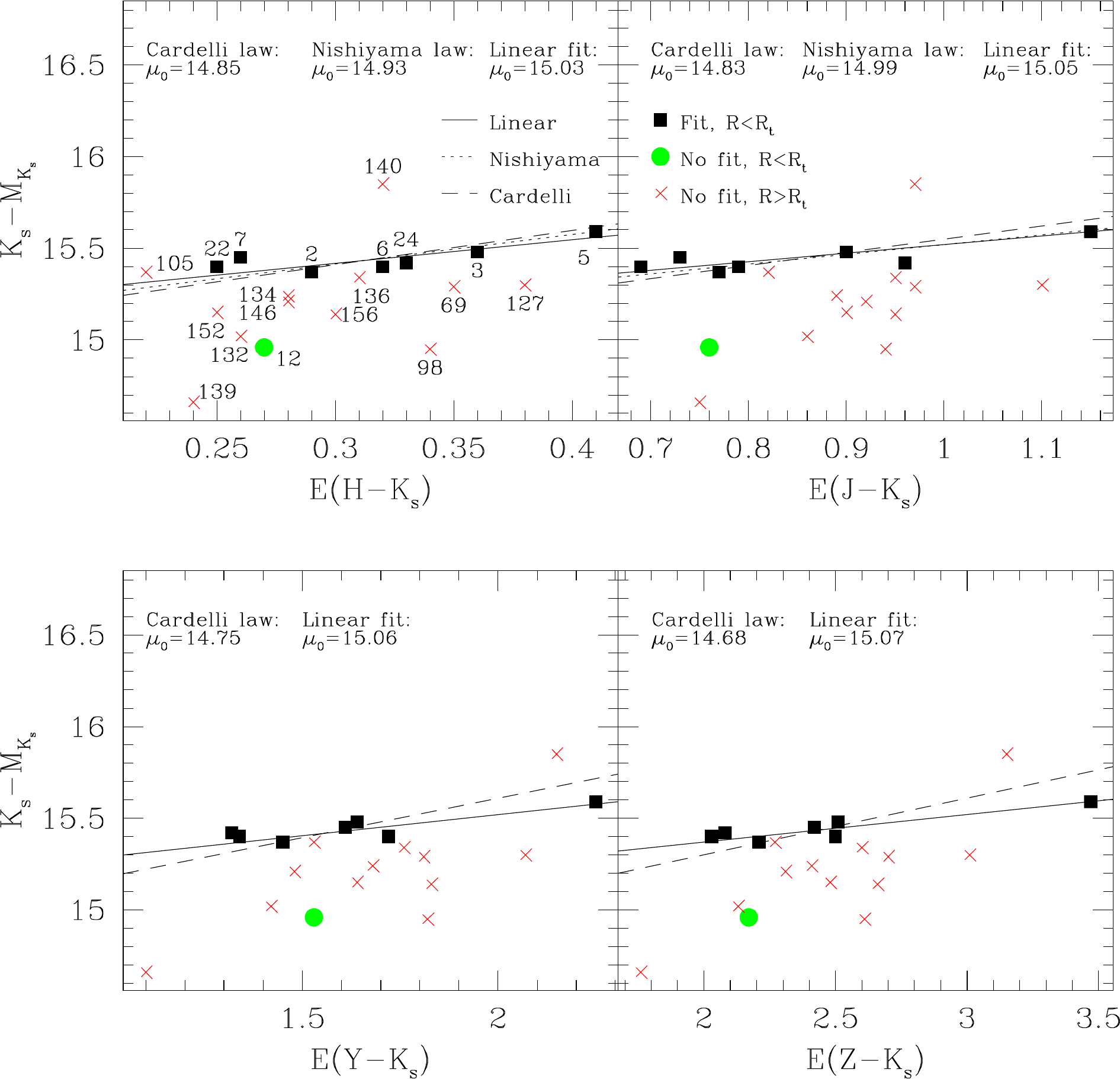}
\caption{As in Figure~\ref{fig_ext2ms02}, but for Terzan~10.}
\label{fig_extt10}
\end{figure*}

As in Section~\ref{sec_dis2ms02}, we did not assume an a priori
extinction law to relate color excess with apparent distance modulus
of the RRab candidates. Instead, using the ordinary least square
bisector method, we got a linear fit between both parameters, which
provide us with values for the extinction ratios
$R_{K_s,\lambda-K_s}$, as the slope of the fit, and an absolute
distance modulus, as the zero-point of the fit (see
Figure~\ref{fig_extt10}). In our analysis, the only clear outlier
among the RR~Lyrae inside the tidal radii, and therefore excluded from
the fit, was NV12 (see discussion in
Section~\ref{sec_vart10in}). Some of the RR~Lyrae in the surrounding
field outside the tidal radius could belong to the cluster according
to their distances, although this is not very plausible (see
discussion in Section~\ref{sec_vart10out}). Again, the ratios
$R_{K_s,\lambda-K_s}$ derived from the fit differ from those of the
standard \citet{car89} law, and agree better with those derived by
\citet{nis09}, as we can see in Table~\ref{tab_ext}. The fact that the
zero-point of the fits, i.e., the absolute distance modulus, show a
much smaller spread among the values derived from using the different
filter-combinations, than if we just use the \citet{car89} law (see
Figure~\ref{fig_extt10}), reinforces our confidence in the fitted
values. We adopt the weighted average of the zero-points from the
different color-excess fits, as the absolute distance modulus of
Terzan~10. We obtain a value $\mu_{0}=15.06\pm0.03\pm0.03$, where
again the first ${\sigma}$ is obtained from the statistical errors in
the fit, while the second one comes from omitting two of the fitted
stars, NV22 and NV5, which have the smallest and biggest extintion
values. This value of $\mu_{0}$ implies the cluster is much farther
away from us than previously thought, almost by a factor of 2 (see
Table~\ref{tab_dis}). This also implies that Terzan~10 is beyond the
Galactic center, in the far side of the Galactic bulge. Until now it
was believed that all the inner Galactic GCs found so far were located
in the near side of the bulge \citep{bar98}.  For the sake of
comparison, a good reference value for the extinction towards
Terzan~10 is $E(J-K_s)=0.86\pm0.16$, the mean of the cluster RRab
candidates that made our fit, but we emphasize that extinction towards
this cluster is also highly variable, with almost 50\% changes over
small regions, and with significant deviations from the standard
extinction law.

The average period of Terzan~10 RRab variable candidates used in the
fit (NV2, NV3, NV5, NV6, NV7, NV22 and NV24) is $\langle P_{ab} \rangle
=0.66\pm0.06$ days, which makes Terzan~10 an Oosterhoff II
cluster. However, its iron-content is significantly higher than the
ones of the other Galactic (and even extragalactic) Oosterhoff II GCs
(see Figure~\ref{fig_oos}). But we should remember here that the
metallicity value adopted in this work is the average of two previous
works (${\rm [Fe/H]}=-0.7$ by \citet{liu94}, and ${\rm [Fe/H]}=-1.2$
by \citet{bic98}), as we mentioned in the beginning of this
section. If ${\rm [Fe/H]}=-0.7$ is closer to the true value, this will
put Terzan~10 closer to NGC~6388 and NGC~6441, in the scarcely
populated Oosterhoff III group of metal-rich clusters with
RR~Lyrae. If, on the other hand, ${\rm [Fe/H]}=-1.2$ is closer to
reality, Terzan~10 would still be the most metal-rich cluster in the
Oosterhoff II group, and would still stand out significantly from the
locus of Galactic GCs. This uncertainty makes a strong case to obtain
spectroscopic metallicity measurements from individual stars in this
GC.

\section{Summary}
\label{sec_sum}
We have shown the potential of the VVV survey for studying the
variable stars of the inner Galactic GCs, by analyzing two
highly-reddened GCs of this sample, 2MASS-GC02 and Terzan~10. We have
discovered 32 new variables inside the tidal radius of 2MASS-GC02 and
70 in a close surrounding region, while for Terzan~10 we have found 48
new variables inside its tidal radius, and 112 in its immediate
surroundings. In both GCs, we have found a significant number of
fundamental-mode RR~Lyrae (12 in 2MASS-GC02, 7 in Terzan~10) that we
have used to accurately measure the extinctions and distances of these
GCs, and to explore the non-standard extinction law in their
directions. Both clusters are closer to the Galactic center than
previously thought ($R_{\rm GC}=1.8$ kpc for 2MASS-GC02; $R_{\rm
  GC}=2.1$ kpc for Terzan~10). We have also found Terzan~10 to be
beyond the Galactic center, making it the only currently known GC to
be on the far side of the Galactic bulge. Extinction towards both
clusters is elevated (especially towards 2MASS-GC02) and highly
differential, and it follows a non-standard law, with values for the
selective-to-total extinction ratios similar to those quoted by
\citet{nis09}.  We have also found both clusters to have quite
uncommon Oosterhoff properties. Terzan~10 can be one of the most
metal-rich Oosterhoff II GCs found in the Galaxy, or belong to the
scarcely-populated metal-rich Oosterhoff III group, depending on the
true metallicity of this cluster. The latter case would imply a
possible He self-enrichment scenario, in analogy to the other two
Oosterhoff III GCs (see \citet{cat09a}, for a review and extensive
references). 2MASS-GC02, on the other hand, is located in the
Oosterhoff gap where very few of the Galactic globular clusters lie,
which may suggest an extragalactic origin for this GC.

There are 34 more known inner Galactic GCs covered by the VVV survey,
and the study of their variable stars, which we plan to present in
future papers in this series, will greatly contribute to better
establish their physical parameters, and to understand the origin of
the Oosterhoff dichotomy in our Galaxy.



\acknowledgments

  We thank Andrzej Udalski for his help comparing some of our detected
  variables with unpublished OGLE data to better determine their
  variability type. This project is supported by the Chilean Ministry
  for the Economy, Development, and Tourism's Programa Iniciativa
  Cient\'ifica Milenio through grant IC120009, awarded to the
  Millennium Institute of Astrophysics (MAS); by Proyecto Fondecyt
  Postdoctoral 3130552 and 3140575; by Proyecto Fondecyt Regular
  1141141; by CONICYT-PCHA Mag\'ister Nacional 2014-22141509; and by
  the ALMA-CONICYT project 31110002.



{\it Facility:} \facility{ESO:VISTA (VIRCAM)}.




\clearpage






\clearpage

\begin{deluxetable*}{ccccccc}
  \tablewidth{0pc} 
  \tablecolumns{10}
  \tablecaption{Physical parameters of the studied clusters, according to the 2010 version of the \citet{har96} catalog. \label{tab_gc}}
  
  \tablehead{\colhead{Cluster} &
    \colhead{$l$} &
    \colhead{$b$} &
    \colhead{${\rm [Fe/H]}$} &
    \colhead{$M_V$} &
    \colhead{$r_{h}$} &
    \colhead{$r_{t}$} \\
     & \colhead{(deg)} & \colhead{(deg)} & & & \colhead{(arcmin)} & \colhead{(arcmin)}} 
  
  \startdata
2MASS-GC02&9.79&-0.61&-1.08&-4.86&0.55&4.90\\
Terzan~10&4.42\tablenotemark{1}&-1.89\tablenotemark{1}&-1.00&-6.35&1.55&5.06\\
  \enddata

  \tablenotetext{1}{Coordinates of Terzan~10 were taken from the 2003 version of \citet{har96} catalog, after analysis of astrometry of VVV images showed that the coordinates provided in the 2010 edition were clearly off.}
\end{deluxetable*}

\begin{deluxetable*}{ccccc}
\tablecaption{Summary of the VVV observations used. \label{tab_obs}}
\tablewidth{0pt}
\tablehead{
\colhead{Cluster} & \colhead{VVV field ID} & 
\colhead{Filter} & \colhead{Exp.Time\tablenotemark{1}} & 
\colhead{Epochs\tablenotemark{2}}
}
\startdata
2MASS-GC02 & b326 & $K_s$ & 8 & 43 (1-11-21-10)\\
& & $H$ & 8 & 1 (1-0-0-0) \\
& & $J$ & 24 & 1 (1-0-0-0) \\
& & $Y$ & 20 & 2 (1-1-0-0) \\
& & $Z$ & 20 & 2 (1-1-0-0) \\
Terzan~10 & b308 & $K_s$ & 8 & 101 (3-7-80-11) \\ 
& & $H$ & 8 & 1 (1-0-0-0) \\
& & $J$ & 24 & 1 (1-0-0-0) \\
& & $Y$ & 20 & 1 (1-0-0-0) \\
& & $Z$ & 20 & 1 (1-0-0-0) \\
\enddata
  \tablenotetext{1}{Effective exposure time, in seconds}
  \tablenotetext{2}{Number of epochs the cluster was observed, with the number of epochs per year from 2010 to 2013 in parentheses}
\end{deluxetable*}

\begin{deluxetable*}{cccccccccccc}
\tablecaption{Properties of the variable candidates in 2MASS-GC02 and immediate surroundings. \label{tab_var2ms02}}
\tablewidth{0pt}
\tablehead{
\colhead{ID\tablenotemark{1}} & \colhead{R.A.(J2000)} & 
\colhead{Dec.(J2000)} & \colhead{$d$\tablenotemark{2}} & 
\colhead{Period} & \colhead{$A_{K_s}$} & \colhead{$\langle K_s \rangle$} & 
\colhead{$Z-K_s$\tablenotemark{3}} & \colhead{$Y-K_s$\tablenotemark{3}} & 
\colhead{$J-K_s$\tablenotemark{3}} & \colhead{$H-K_s$\tablenotemark{3}} & \colhead{Type}\\
& \colhead{h:m:s} & \colhead{d:m:s} & \colhead{arcmin} & \colhead{days} & & & & & & & 
}
\startdata
NV1  & 18:09:35.98 & -20:47:11.6 & 0.47 & 0.700046 & 0.320 & 14.566 & \nodata & 4.313 & 2.699 & 0.956 & RRab \\
NV2  & 18:09:34.22 & -20:46:56.1 & 0.57 & 0.651668 & 0.232 & 14.754 & \nodata & \nodata & 2.881 & 0.997 & RRab \\
NV3  & 18:09:33.77 & -20:46:29.4 & 0.68 & 0.570430 & 0.266 & 15.040 & \nodata & \nodata & 3.199 & 1.094 & RRab \\
NV4  & 18:09:38.59 & -20:46:04.9 & 0.81 & 0.623735 & 0.233 & 15.301 & \nodata & \nodata & 4.573 & 1.488 & RRab \\
NV5  & 18:09:38.86 & -20:47:26.4 & 0.90 & 0.603298 & 0.312 & 14.950 & \nodata & \nodata & 3.257 & 1.089 & RRab \\
NV6  & 18:09:33.00 & -20:47:07.7 & 0.91 & 0.551331 & 0.386 & 14.943 & \nodata & \nodata & 2.941 & 0.989 & RRab \\
NV7  & 18:09:29.99 & -20:45:48.2 & 1.78 & 0.569916 & 0.294 & 15.453 & \nodata & \nodata & 3.736 & 1.337 & RRab \\
NV8  & 18:09:37.70 & -20:48:37.5 & 1.91 & 0.580325 & 0.269 & 14.817 & \nodata & \nodata & 2.867 & 0.969 & RRab \\
NV9  & 18:09:45.01 & -20:46:09.1 & 2.07 & 0.608760 & 0.256 & 15.079 & \nodata & \nodata & 3.659 & 1.218 & RRab \\
NV10 & 18:09:29.23 & -20:48:10.9 & 2.23 & 0.489329 & 0.312 & 15.217 & \nodata & \nodata & 3.395 & 1.217 & RRab \\
\enddata
\tablecomments{Table~\ref{tab_var2ms02} is published in its entirety in the 
electronic edition of the {\it Astronomical Journal}.  A portion is 
shown here for guidance regarding its form and content.}
\tablenotetext{1}{Asterisk means that the physical parameters of the star should be considered with caution, since the light curve is not well defined yet.}
\tablenotetext{2}{Distance to the cluster center}
\tablenotetext{3}{Colors measured from single-epoch measurements in $Z$,$Y$,$J$, and $H$ (see Section~\ref{sec_dis2ms02} for further explanation).}  
\end{deluxetable*}

\clearpage

\begin{deluxetable*}{ccccc}
  \tablewidth{0pc} 
  \tablecaption{Selective-to-total extinction ratios towards the studied GCs\label{tab_ext}}
  
  \tablehead{  &
    \colhead{2MASS-GC02} &
    \colhead{Terzan~10} & 
    \colhead{Nishiyama09} & 
    \colhead{Cardelli89}} 
  
  \startdata
$A_{K_s}/E(H-K_s)$ & 1.27$\pm$0.18$\pm$0.23 & 1.29$\pm$0.23$\pm$0.3 & 1.61$\pm$0.04 & 1.87\\
$A_{K_s}/E(J-K_s)$ & 0.40$\pm$0.08$\pm$0.13 & 0.47$\pm$0.05$_{+0.03}^{+0.03}$ & 0.528$\pm$0.015 & 0.72\\
$A_{K_s}/E(Y-K_s)$ & \nodata & 0.23$\pm$0.02$_{+0.10}^{+0.02}$ & \nodata & 0.43 \\
$A_{K_s}/E(Z-K_s)$ & \nodata & 0.15$\pm$0.02$_{+0.07}^{+0.01}$ & \nodata & 0.31 \\

  \enddata
\end{deluxetable*}

\begin{deluxetable*}{ccccccc}
  \tablewidth{0pc} 
  \tablecaption{Distances and extinctions to the studied clusters\label{tab_dis}}
  
  \tablehead{  & 
    \colhead{$R_{\odot,\rm Harris96}$} &
    \colhead{$R_{GC,\rm Harris96}$\tablenotemark{1}} &
    \colhead{$E(B-V)_{\rm Harris96}$} &
    \colhead{$R_{\odot,\rm derived}$} &
    \colhead{$R_{GC,\rm derived}$\tablenotemark{1}} &
    \colhead{$E(J-K_s)_{\rm derived}$} \\
     & \colhead{(kpc)} & \colhead{(kpc)} & & \colhead{(kpc)} & \colhead{(kpc)} & } 
  
  \startdata
2MASS-GC02&4.9&$-3.6$&5.16&7.1$\pm$0.5$\pm$0.9&$-1.8\pm$0.3$\pm$0.6&3.1\\
Terzan~10&5.8&$-2.6$&2.40&10.3$\pm$0.2$\pm$0.2&$+2.1\pm$0.2$\pm$0.2&0.86\\
  \enddata
  \tablenotetext{1}{The minus sign indicates location on the near side of the bulge, and the plus sign indicates location on the far side of the bulge. Distances obtained assuming a Galactocentric distance for the Sun of 8.3 kpc \citep{gil09,dek13}}
\end{deluxetable*}

\begin{deluxetable*}{cccccccccccc}
\tablecaption{Properties of the variable candidates in Terzan~10 and immediate surroundings. \label{tab_vart10}}
\tablewidth{0pt}
\tablehead{
\colhead{ID\tablenotemark{1}} & \colhead{R.A.(J2000)} & 
\colhead{Dec.(J2000)} & \colhead{$d$\tablenotemark{2}} & 
\colhead{Period} & \colhead{$A_{K_s}$} & \colhead{$\langle K_s \rangle$} & 
\colhead{$Z-K_s$\tablenotemark{3}} & \colhead{$Y-K_s$\tablenotemark{3}} & 
\colhead{$J-K_s$\tablenotemark{3}} & \colhead{$H-K_s$\tablenotemark{3}} & \colhead{Type}\\
& \colhead{h:m:s} & \colhead{d:m:s} & \colhead{arcmin} & \colhead{days} & & & & & & & 
}
\startdata
NV1   & 18:02:58.87& -26:03:35.2& 0.53& 3.8798    &0.402 &14.072 &2.784 &1.942 &1.136 &0.377 & Ecl \\
NV2   & 18:02:59.48& -26:04:22.5& 0.6 & 0.730516  &0.364 &14.608 &2.789 &1.916 &1.068 &0.364 & RRab\\
NV3   & 18:02:54.05& -26:03:46.9& 0.78& 0.701719  &0.323 &14.759 &3.073 &2.092 &1.186 &0.426 & RRab\\
NV4   & 18:03:00.19& -26:05:05.8& 1.26& 0.684468  &0.238 &14.545 &2.241 &1.593 &0.84  &0.305 & Ecl \\
NV5   & 18:02:57.14& -26:02:43.9& 1.27& 0.688512  &0.294 &14.889 &4.023 &2.695 &1.43  &0.481 & RRab\\
NV6   & 18:02:56.90& -26:05:19.6& 1.33& 0.582339  &0.322 &14.875 &2.97  &2.105 &1.03  &0.384 & RRab\\
NV7   & 18:02:53.23& -26:05:12.7& 1.53& 0.715284  &0.33  &14.711 &2.992 &2.068 &1.016 &0.328 & RRab\\
NV8   & 18:03:03.84& -26:03:13.4& 1.64& 1.06527   &0.295 &16.779 &2.674 &\nodata    &0.957 &0.149 & \nodata\\
NV9   & 18:02:51.23& -26:05:14.8& 1.87& 0.193833  &0.259 &16.091 &2.54  &1.82  &0.989 &0.277 & \nodata\\
NV10  & 18:03:04.64& -26:05:15.4& 2.05& 0.211256  &0.41  &16.203 &2.038 &1.376 &0.816 &0.267 & \nodata\\
\enddata
\tablecomments{Table~\ref{tab_vart10} is published in its entirety in the 
electronic edition of the {\it Astronomical Journal}.  A portion is 
shown here for guidance regarding its form and content.}
\tablenotetext{1}{Asterisk means that the physical parameters of the star should be considered with caution, since the light curve is not well defined yet.}
\tablenotetext{2}{Distance to the cluster center}
\tablenotetext{3}{Colors measured from single-epoch measurements in $Z$,$Y$,$J$, and $H$ (see Section~\ref{sec_dist10} for further explanation).}  
\end{deluxetable*}



\begin{figure*}
\plotone{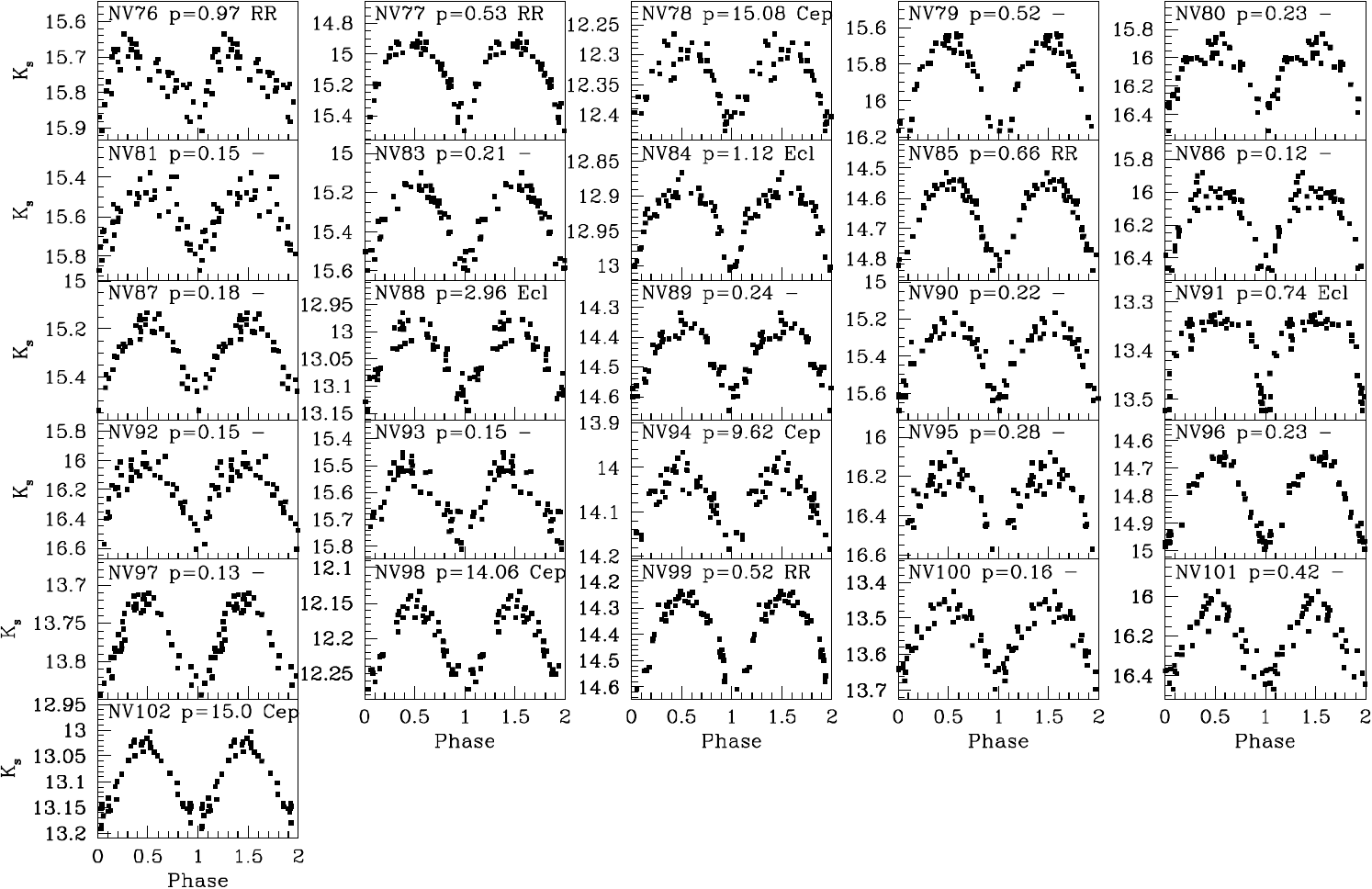}
\end{figure*}
\begin{figure*}
\plotone{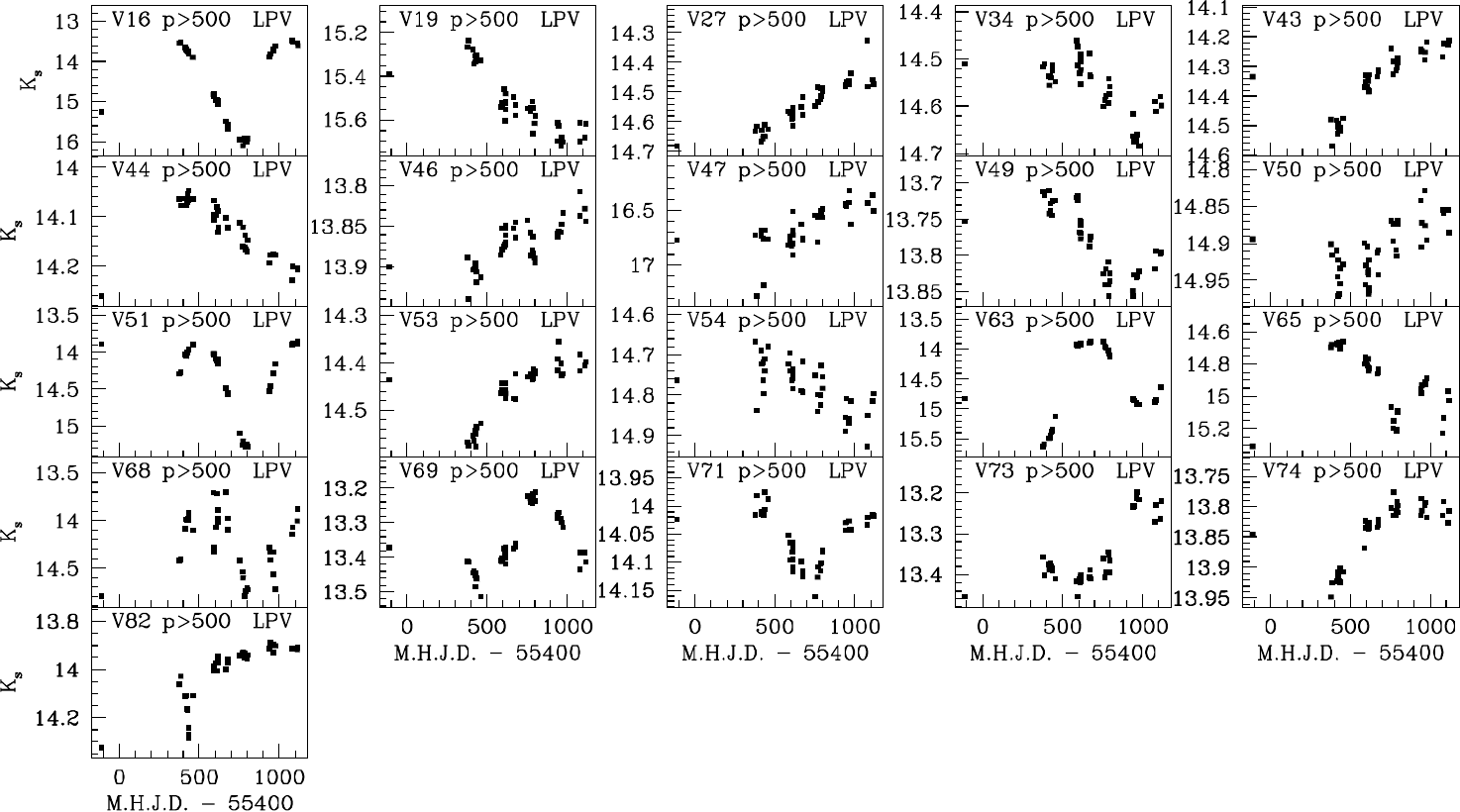}
\end{figure*} 

\begin{figure*}
\plotone{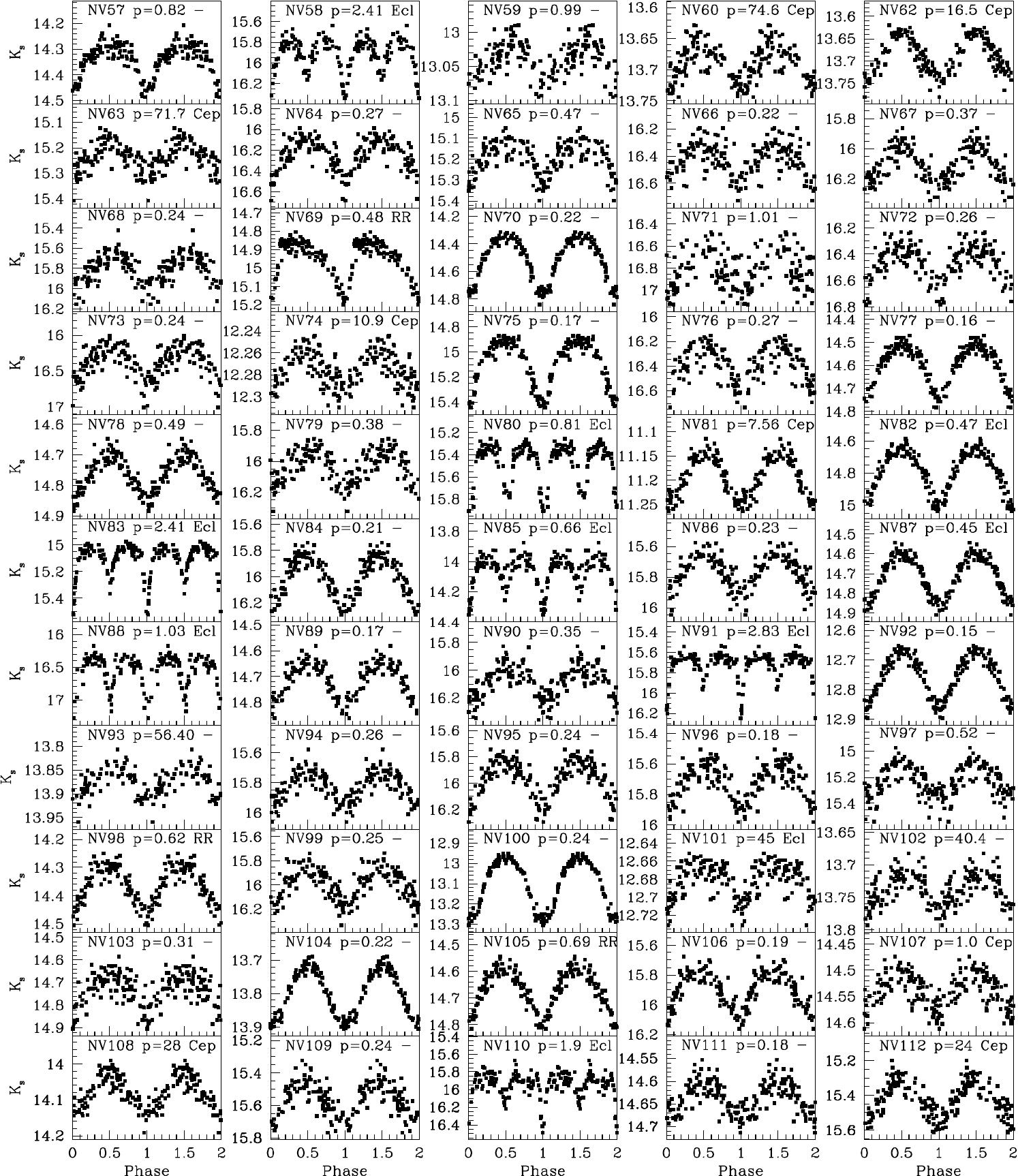}
\end{figure*}
\begin{figure*}
\plotone{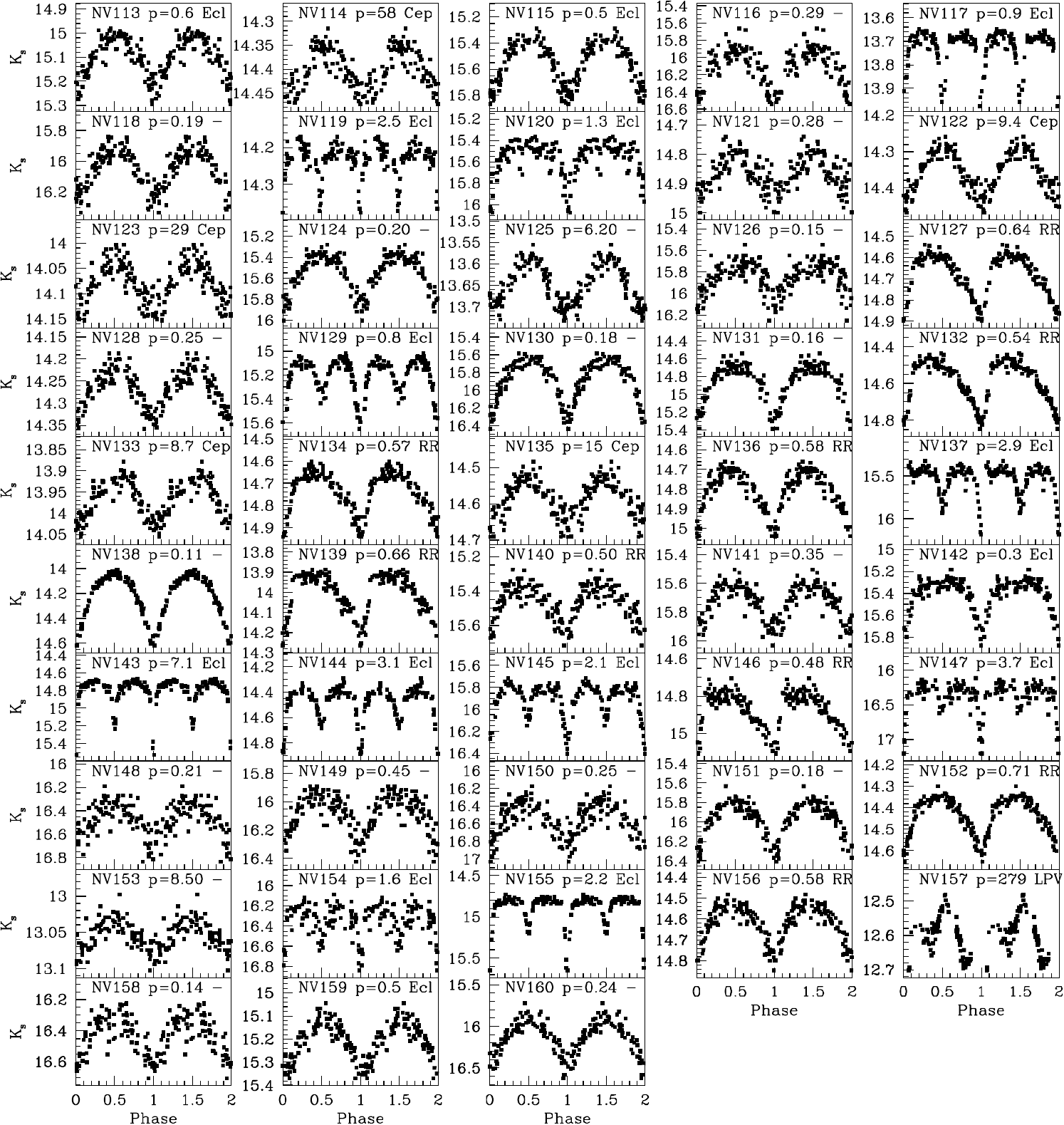}
\end{figure*}
\begin{figure*}
\epsscale{0.45}
\plotone{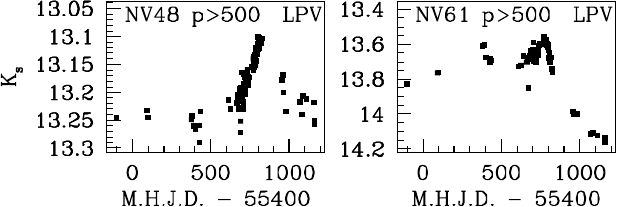}
\epsscale{1}
\end{figure*}

\begin{thebibliography}{}
\expandafter\ifx\csname natexlab\endcsname\relax\def\natexlab#1{#1}\fi

\bibitem[{{Alonso-Garc{\'{\i}}a} {et~al.}(2012){Alonso-Garc{\'{\i}}a}, {Mateo},
  {Sen}, {Banerjee}, {Catelan}, {Minniti}, \& {von Braun}}]{alo12}
{Alonso-Garc{\'{\i}}a}, J., {Mateo}, M., {Sen}, B., {et~al.} 2012, \aj, 143, 70

\bibitem[{{Angeloni} {et~al.}(2014){Angeloni}, {Contreras Ramos}, {Catelan},
  {D{\'e}k{\'a}ny}, {Gran}, {Alonso-Garc{\'{\i}}a}, {Hempel}, {Navarrete},
  {Andrews}, {Aparicio}, {Beam{\'{\i}}n}, {Berger}, {Borissova}, {Contreras
  Pe{\~n}a}, {Cunial}, {de Grijs}, {Espinoza}, {Eyheramendy}, {Ferreira Lopes},
  {Fiaschi}, {Hajdu}, {Han}, {He{\l}miniak}, {Hempel}, {Hidalgo}, {Ita},
  {Jeon}, {Jord{\'a}n}, {Kwon}, {Lee}, {Mart{\'{\i}}n}, {Masetti}, {Matsunaga},
  {Milone}, {Minniti}, {Morelli}, {Murgas}, {Nagayama}, {Navarro}, {Ochner},
  {P{\'e}rez}, {Pichara}, {Rojas-Arriagada}, {Roquette}, {Saito}, {Siviero},
  {Sohn}, {Sung}, {Tamura}, {Tata}, {Tomasella}, {Townsend}, \&
  {Whitelock}}]{ang14}
{Angeloni}, R., {Contreras Ramos}, R., {Catelan}, M., {et~al.} 2014, \aap, 567,
  A100

\bibitem[{{Barbuy} {et~al.}(1998){Barbuy}, {Bica}, \& {Ortolani}}]{bar98}
{Barbuy}, B., {Bica}, E., \& {Ortolani}, S. 1998, \aap, 333, 117

\bibitem[{{Bica} {et~al.}(1998){Bica}, {Claria}, {Piatti}, \&
  {Bonatto}}]{bic98}
{Bica}, E., {Claria}, J.~J., {Piatti}, A.~E., \& {Bonatto}, C. 1998, \aaps,
  131, 483

\bibitem[{{Bono} {et~al.}(2001){Bono}, {Caputo}, {Castellani}, {Marconi}, \&
  {Storm}}]{bon01}
{Bono}, G., {Caputo}, F., {Castellani}, V., {Marconi}, M., \& {Storm}, J. 2001,
  \mnras, 326, 1183

\bibitem[{{Borissova} {et~al.}(2007){Borissova}, {Ivanov}, {Stephens},
  {Catelan}, {Minniti}, \& {Prieto}}]{bor07}
{Borissova}, J., {Ivanov}, V.~D., {Stephens}, A.~W., {et~al.} 2007, \aap, 474,
  121

\bibitem[{{Borissova} {et~al.}(2002){Borissova}, {Ivanov}, \& {Vanzi}}]{bor02}
{Borissova}, J., {Ivanov}, V.~D., \& {Vanzi}, L. 2002, in IAU Symposium, Vol.
  207, Extragalactic Star Clusters, ed. D.~P. {Geisler}, E.~K. {Grebel}, \&
  D.~{Minniti}, 107

\bibitem[{{Caloi} \& {D'Antona}(2007)}]{cal07}
{Caloi}, V., \& {D'Antona}, F. 2007, \aap, 463, 949

\bibitem[{{Cardelli} {et~al.}(1989){Cardelli}, {Clayton}, \& {Mathis}}]{car89}
{Cardelli}, J.~A., {Clayton}, G.~C., \& {Mathis}, J.~S. 1989, \apj, 345, 245

\bibitem[{{Cassisi} {et~al.}(2004){Cassisi}, {Castellani}, {Caputo}, \&
  {Castellani}}]{cas04}
{Cassisi}, S., {Castellani}, M., {Caputo}, F., \& {Castellani}, V. 2004, \aap,
  426, 641

\bibitem[{{Catelan}(2009{\natexlab{a}})}]{cat09a}
{Catelan}, M. 2009{\natexlab{a}}, \apss, 320, 261

\bibitem[{{Catelan}(2009{\natexlab{b}})}]{cat09b}
{Catelan}, M. 2009{\natexlab{b}}, in IAU Symposium, Vol. 258, IAU Symposium,
  ed. E.~E. {Mamajek}, D.~R. {Soderblom}, \& R.~F.~G. {Wyse}, 209--220

\bibitem[{{Catelan} {et~al.}(2004){Catelan}, {Pritzl}, \& {Smith}}]{cat04}
{Catelan}, M., {Pritzl}, B.~J., \& {Smith}, H.~A. 2004, \apjs, 154, 633

\bibitem[{{Catelan} {et~al.}(2011){Catelan}, {Minniti}, {Lucas},
  {Alonso-Garc{\'{\i}}a}, {Angeloni}, {Beam{\'{\i}}n}, {Bonatto}, {Borissova},
  {Contreras}, {Cross}, {D{\'e}k{\'a}any}, {Emerson}, {Eyheramendy}, {Geisler},
  {Gonz{\'a}lez-Solares}, {Helminiak}, {Hempel}, {Irwin}, {Ivanov},
  {Jord{\'a}n}, {Kerins}, {Kurtev}, {Mauro}, {Moni Bidin}, {Navarrete},
  {P{\'e}rez}, {Pichara}, {Read}, {Rejkuba}, {Saito}, {Sale}, \&
  {Toledo}}]{cat11}
{Catelan}, M., {Minniti}, D., {Lucas}, P.~W., {et~al.} 2011, in RR Lyrae Stars,
  Metal-Poor Stars, and the Galaxy, ed. A.~{McWilliam}, 145

\bibitem[{{Catelan} {et~al.}(2013){Catelan}, {Minniti}, {Lucas},
  {D{\'e}k{\'a}ny}, {Saito}, {Angeloni}, {Alonso-Garc{\'{\i}}a}, {Hempel},
  {Helminiak}, {Jord{\'a}n}, {Contreras Ramos}, {Navarrete}, {Beam{\'{\i}}n},
  {Rojas}, {Gran}, {Ferreira Lopes}, {Contreras Pe{\~n}a}, {Kerins},
  {Huckvale}, {Rejkuba}, {Cohen}, {Mauro}, {Borissova}, {Amigo}, {Eyheramendy},
  {Pichara}, {Espinoza}, {Navarro}, {Hajdu}, {Calder{\'o}n Espinoza}, {Muro},
  {Andrews}, {Motta}, {Kurtev}, {Emerson}, {Moni Bidin}, \&
  {Chen{\'e}}}]{cat13}
{Catelan}, M., {Minniti}, D., {Lucas}, P.~W., {et~al.} 2013, ArXiv e-prints,
  arXiv:1310.1996

\bibitem[{{Clement} {et~al.}(2001){Clement}, {Muzzin}, {Dufton}, {Ponnampalam},
  {Wang}, {Burford}, {Richardson}, {Rosebery}, {Rowe}, \& {Hogg}}]{cle01}
{Clement}, C.~M., {Muzzin}, A., {Dufton}, Q., {et~al.} 2001, \aj, 122, 2587

\bibitem[{{Corwin} {et~al.}(2006){Corwin}, {Sumerel}, {Pritzl}, {Smith},
  {Catelan}, {Sweigart}, \& {Stetson}}]{cor06}
{Corwin}, T.~M., {Sumerel}, A.~N., {Pritzl}, B.~J., {et~al.} 2006, \aj, 132,
  1014

\bibitem[{{D{\'e}k{\'a}ny} {et~al.}(2013){D{\'e}k{\'a}ny}, {Minniti},
  {Catelan}, {Zoccali}, {Saito}, {Hempel}, \& {Gonzalez}}]{dek13}
{D{\'e}k{\'a}ny}, I., {Minniti}, D., {Catelan}, M., {et~al.} 2013, \apjl, 776,
  L19

\bibitem[{{Emerson} {et~al.}(2004){Emerson}, {Irwin}, {Lewis}, {Hodgkin},
  {Evans}, {Bunclark}, {McMahon}, {Hambly}, {Mann}, {Bond}, {Sutorius}, {Read},
  {Williams}, {Lawrence}, \& {Stewart}}]{eme04}
{Emerson}, J.~P., {Irwin}, M.~J., {Lewis}, J., {et~al.} 2004, in Society of
  Photo-Optical Instrumentation Engineers (SPIE) Conference Series, Vol. 5493,
  Optimizing Scientific Return for Astronomy through Information Technologies,
  ed. P.~J. {Quinn} \& A.~{Bridger}, 401--410

\bibitem[{{Feast} {et~al.}(2008){Feast}, {Laney}, {Kinman}, {van Leeuwen}, \&
  {Whitelock}}]{fea08}
{Feast}, M.~W., {Laney}, C.~D., {Kinman}, T.~D., {van Leeuwen}, F., \&
  {Whitelock}, P.~A. 2008, \mnras, 386, 2115

\bibitem[{{Gillessen} {et~al.}(2009){Gillessen}, {Eisenhauer}, {Trippe},
  {Alexander}, {Genzel}, {Martins}, \& {Ott}}]{gil09}
{Gillessen}, S., {Eisenhauer}, F., {Trippe}, S., {et~al.} 2009, \apj, 692, 1075

\bibitem[{{Hambly} {et~al.}(2004){Hambly}, {Mann}, {Bond}, {Sutorius}, {Read},
  {Williams}, {Lawrence}, \& {Emerson}}]{ham04}
{Hambly}, N.~C., {Mann}, R.~G., {Bond}, I., {et~al.} 2004, in Society of
  Photo-Optical Instrumentation Engineers (SPIE) Conference Series, Vol. 5493,
  Optimizing Scientific Return for Astronomy through Information Technologies,
  ed. P.~J. {Quinn} \& A.~{Bridger}, 423--431

\bibitem[{{Harris}(1996)}]{har96}
{Harris}, W.~E. 1996, \aj, 112, 1487

\bibitem[{{Hurt} {et~al.}(2000){Hurt}, {Jarrett}, {Kirkpatrick}, {Cutri},
  {Schneider}, {Skrutskie}, \& {van Driel}}]{hur00}
{Hurt}, R.~L., {Jarrett}, T.~H., {Kirkpatrick}, J.~D., {et~al.} 2000, \aj, 120,
  1876

\bibitem[{{Irwin} {et~al.}(2004){Irwin}, {Lewis}, {Hodgkin}, {Bunclark},
  {Evans}, {McMahon}, {Emerson}, {Stewart}, \& {Beard}}]{irw04}
{Irwin}, M.~J., {Lewis}, J., {Hodgkin}, S., {et~al.} 2004, in Society of
  Photo-Optical Instrumentation Engineers (SPIE) Conference Series, Vol. 5493,
  Optimizing Scientific Return for Astronomy through Information Technologies,
  ed. P.~J. {Quinn} \& A.~{Bridger}, 411--422

\bibitem[{{Isobe} {et~al.}(1990){Isobe}, {Feigelson}, {Akritas}, \&
  {Babu}}]{iso90}
{Isobe}, T., {Feigelson}, E.~D., {Akritas}, M.~G., \& {Babu}, G.~J. 1990, \apj,
  364, 104

\bibitem[{{Ivanov} {et~al.}(2000){Ivanov}, {Borissova}, \& {Vanzi}}]{iva00}
{Ivanov}, V.~D., {Borissova}, J., \& {Vanzi}, L. 2000, \aap, 362, L1

\bibitem[{{Kov{\'a}cs} {et~al.}(2005){Kov{\'a}cs}, {Bakos}, \& {Noyes}}]{kov05}
{Kov{\'a}cs}, G., {Bakos}, G., \& {Noyes}, R.~W. 2005, \mnras, 356, 557

\bibitem[{{Kunder} {et~al.}(2008){Kunder}, {Popowski}, {Cook}, \&
  {Chaboyer}}]{kun08}
{Kunder}, A., {Popowski}, P., {Cook}, K.~H., \& {Chaboyer}, B. 2008, \aj, 135,
  631

\bibitem[{{Liu} {et~al.}(1994){Liu}, {McLean}, \& {Becklin}}]{liu94}
{Liu}, T., {McLean}, I., \& {Becklin}, E. 1994, in Astrophysics and Space
  Science Library, Vol. 190, Astronomy with Arrays, The Next Generation, ed.
  I.~S. {McLean}, 101

\bibitem[{{Longmore} {et~al.}(1990){Longmore}, {Dixon}, {Skillen}, {Jameson},
  \& {Fernley}}]{lon90}
{Longmore}, A.~J., {Dixon}, R., {Skillen}, I., {Jameson}, R.~F., \& {Fernley},
  J.~A. 1990, \mnras, 247, 684

\bibitem[{{Longmore} {et~al.}(1986){Longmore}, {Fernley}, \& {Jameson}}]{lon86}
{Longmore}, A.~J., {Fernley}, J.~A., \& {Jameson}, R.~F. 1986, \mnras, 220, 279

\bibitem[{{Matsunaga} {et~al.}(2013){Matsunaga}, {Feast}, {Kawadu},
  {Nishiyama}, {Nagayama}, {Nagata}, {Tamura}, {Bono}, \& {Kobayashi}}]{mat13}
{Matsunaga}, N., {Feast}, M.~W., {Kawadu}, T., {et~al.} 2013, \mnras, 429, 385

\bibitem[{{Minniti} {et~al.}(2010){Minniti}, {Lucas}, {Emerson}, {Saito},
  {Hempel}, {Pietrukowicz}, {Ahumada}, {Alonso}, {Alonso-Garcia}, {Arias},
  {Bandyopadhyay}, {Barb{\'a}}, {Barbuy}, {Bedin}, {Bica}, {Borissova},
  {Bronfman}, {Carraro}, {Catelan}, {Clari{\'a}}, {Cross}, {de Grijs},
  {D{\'e}k{\'a}ny}, {Drew}, {Fari{\~n}a}, {Feinstein}, {Fern{\'a}ndez
  Laj{\'u}s}, {Gamen}, {Geisler}, {Gieren}, {Goldman}, {Gonzalez}, {Gunthardt},
  {Gurovich}, {Hambly}, {Irwin}, {Ivanov}, {Jord{\'a}n}, {Kerins}, {Kinemuchi},
  {Kurtev}, {L{\'o}pez-Corredoira}, {Maccarone}, {Masetti}, {Merlo},
  {Messineo}, {Mirabel}, {Monaco}, {Morelli}, {Padilla}, {Palma}, {Parisi},
  {Pignata}, {Rejkuba}, {Roman-Lopes}, {Sale}, {Schreiber}, {Schr{\"o}der},
  {Smith}, {Sodr{\'e}}, {Soto}, {Tamura}, {Tappert}, {Thompson}, {Toledo},
  {Zoccali}, \& {Pietrzynski}}]{min10}
{Minniti}, D., {Lucas}, P.~W., {Emerson}, J.~P., {et~al.} 2010, New Astronomy,
  15, 433

\bibitem[{{Minniti} {et~al.}(2011){Minniti}, {Hempel}, {Toledo}, {Ivanov},
  {Alonso-Garc{\'{\i}}a}, {Saito}, {Catelan}, {Geisler}, {Jord{\'a}n},
  {Borissova}, {Zoccali}, {Kurtev}, {Carraro}, {Barbuy}, {Clari{\'a}},
  {Rejkuba}, {Emerson}, \& {Moni Bidin}}]{min11b}
{Minniti}, D., {Hempel}, M., {Toledo}, I., {et~al.} 2011, \aap, 527, A81

\bibitem[{{Moni Bidin} {et~al.}(2011){Moni Bidin}, {Mauro}, {Geisler},
  {Minniti}, {Catelan}, {Hempel}, {Valenti}, {Valcarce},
  {Alonso-Garc{\'{\i}}a}, {Borissova}, {Carraro}, {Lucas}, {Chen{\'e}},
  {Zoccali}, \& {Kurtev}}]{mon11}
{Moni Bidin}, C., {Mauro}, F., {Geisler}, D., {et~al.} 2011, \aap, 535, A33

\bibitem[{{Nishiyama} {et~al.}(2009){Nishiyama}, {Tamura}, {Hatano}, {Kato},
  {Tanab{\'e}}, {Sugitani}, \& {Nagata}}]{nis09}
{Nishiyama}, S., {Tamura}, M., {Hatano}, H., {et~al.} 2009, \apj, 696, 1407

\bibitem[{{Nishiyama} {et~al.}(2006){Nishiyama}, {Nagata}, {Kusakabe},
  {Matsunaga}, {Naoi}, {Kato}, {Nagashima}, {Sugitani}, {Tamura}, {Tanab{\'e}},
  \& {Sato}}]{nis06}
{Nishiyama}, S., {Nagata}, T., {Kusakabe}, N., {et~al.} 2006, \apj, 638, 839

\bibitem[{{Oosterhoff}(1939)}]{oos39}
{Oosterhoff}, P.~T. 1939, The Observatory, 62, 104

\bibitem[{{Ortolani} {et~al.}(1997){Ortolani}, {Bica}, \& {Barbuy}}]{ort97}
{Ortolani}, S., {Bica}, E., \& {Barbuy}, B. 1997, \aaps, 126, 319

\bibitem[{{Pritzl} {et~al.}(2000){Pritzl}, {Smith}, {Catelan}, \&
  {Sweigart}}]{pri00}
{Pritzl}, B., {Smith}, H.~A., {Catelan}, M., \& {Sweigart}, A.~V. 2000, \apjl,
  530, L41

\bibitem[{{Pritzl} {et~al.}(2001){Pritzl}, {Smith}, {Catelan}, \&
  {Sweigart}}]{pri01}
{Pritzl}, B.~J., {Smith}, H.~A., {Catelan}, M., \& {Sweigart}, A.~V. 2001, \aj,
  122, 2600

\bibitem[{{Pritzl} {et~al.}(2002){Pritzl}, {Smith}, {Catelan}, \&
  {Sweigart}}]{pri02}
---. 2002, \aj, 124, 949

\bibitem[{{Pritzl} {et~al.}(2003){Pritzl}, {Smith}, {Stetson}, {Catelan},
  {Sweigart}, {Layden}, \& {Rich}}]{pri03}
{Pritzl}, B.~J., {Smith}, H.~A., {Stetson}, P.~B., {et~al.} 2003, \aj, 126,
  1381

\bibitem[{{Pritzl} {et~al.}(2005){Pritzl}, {Venn}, \& {Irwin}}]{pri05}
{Pritzl}, B.~J., {Venn}, K.~A., \& {Irwin}, M. 2005, \aj, 130, 2140

\bibitem[{{Rieke} \& {Lebofsky}(1985)}]{rie85}
{Rieke}, G.~H., \& {Lebofsky}, M.~J. 1985, \apj, 288, 618

\bibitem[{{Saito} {et~al.}(2011){Saito}, {Minniti}, {D{\'e}k{\'a}ny}, {Hempel},
  {Alonso-Garc{\'{\i}}a}, {Toledo}, {Beamin}, {Angeloni}, {Lucas}, \&
  {Emerson}}]{sai11}
{Saito}, R.~K., {Minniti}, D., {D{\'e}k{\'a}ny}, I., {et~al.} 2011, in Revista
  Mexicana de Astronomia y Astrofisica Conference Series, Vol.~40, Revista
  Mexicana de Astronomia y Astrofisica Conference Series, 221--224

\bibitem[{{Saito} {et~al.}(2012){Saito}, {Hempel}, {Minniti}, {Lucas},
  {Rejkuba}, {Toledo}, {Gonzalez}, {Alonso-Garc{\'{\i}}a}, {Irwin},
  {Gonzalez-Solares}, {Hodgkin}, {Lewis}, {Cross}, {Ivanov}, {Kerins},
  {Emerson}, {Soto}, {Am{\^o}res}, {Gurovich}, {D{\'e}k{\'a}ny}, {Angeloni},
  {Beamin}, {Catelan}, {Padilla}, {Zoccali}, {Pietrukowicz}, {Moni Bidin},
  {Mauro}, {Geisler}, {Folkes}, {Sale}, {Borissova}, {Kurtev}, {Ahumada},
  {Alonso}, {Adamson}, {Arias}, {Bandyopadhyay}, {Barb{\'a}}, {Barbuy},
  {Baume}, {Bedin}, {Bellini}, {Benjamin}, {Bica}, {Bonatto}, {Bronfman},
  {Carraro}, {Chen{\`e}}, {Clari{\'a}}, {Clarke}, {Contreras}, {Corvill{\'o}n},
  {de Grijs}, {Dias}, {Drew}, {Fari{\~n}a}, {Feinstein},
  {Fern{\'a}ndez-Laj{\'u}s}, {Gamen}, {Gieren}, {Goldman},
  {Gonz{\'a}lez-Fern{\'a}ndez}, {Grand}, {Gunthardt}, {Hambly}, {Hanson},
  {He{\l}miniak}, {Hoare}, {Huckvale}, {Jord{\'a}n}, {Kinemuchi}, {Longmore},
  {L{\'o}pez-Corredoira}, {Maccarone}, {Majaess}, {Mart{\'{\i}}n}, {Masetti},
  {Mennickent}, {Mirabel}, {Monaco}, {Morelli}, {Motta}, {Palma}, {Parisi},
  {Parker}, {Pe{\~n}aloza}, {Pietrzy{\'n}ski}, {Pignata}, {Popescu}, {Read},
  {Rojas}, {Roman-Lopes}, {Ruiz}, {Saviane}, {Schreiber}, {Schr{\"o}der},
  {Sharma}, {Smith}, {Sodr{\'e}}, {Stead}, {Stephens}, {Tamura}, {Tappert},
  {Thompson}, {Valenti}, {Vanzi}, {Walton}, {Weidmann}, \& {Zijlstra}}]{sai12b}
{Saito}, R.~K., {Hempel}, M., {Minniti}, D., {et~al.} 2012, \aap, 537, A107

\bibitem[{{Schechter} {et~al.}(1993){Schechter}, {Mateo}, \& {Saha}}]{sch93}
{Schechter}, P.~L., {Mateo}, M., \& {Saha}, A. 1993, \pasp, 105, 1342

\bibitem[{{Smith} {et~al.}(2011){Smith}, {Catelan}, \& {Kuehn}}]{smi11}
{Smith}, H.~A., {Catelan}, M., \& {Kuehn}, C. 2011, in RR Lyrae Stars,
  Metal-Poor Stars, and the Galaxy, ed. A.~{McWilliam}, 17

\bibitem[{{Soszy{\'n}ski} {et~al.}(2014){Soszy{\'n}ski}, {Udalski},
  {Szyma{\'n}ski}, {Pietrukowicz}, {Mr{\'o}z}, {Skowron}, {Koz{\l}owski},
  {Poleski}, {Skowron}, {Pietrzy{\'n}ski}, {Wyrzykowski}, {Ulaczyk}, \&
  {Kubiak}}]{sos14}
{Soszy{\'n}ski}, I., {Udalski}, A., {Szyma{\'n}ski}, M.~K., {et~al.} 2014,
  Acta Astron., 64, 177

\bibitem[{{Stellingwerf}(1978)}]{ste78}
{Stellingwerf}, R.~F. 1978, \apj, 224, 953

\bibitem[{{Stetson}(1996)}]{ste96}
{Stetson}, P.~B. 1996, \pasp, 108, 851

\bibitem[{{Taylor}(2006)}]{tay06}
{Taylor}, M.~B. 2006, in Astronomical Society of the Pacific Conference Series,
  Vol. 351, Astronomical Data Analysis Software and Systems XV, ed.
  C.~{Gabriel}, C.~{Arviset}, D.~{Ponz}, \& S.~{Enrique}, 666

\bibitem[{{Terzan}(1971)}]{ter71}
{Terzan}, A. 1971, \aap, 12, 477

\bibitem[{{Valenti} {et~al.}(2007){Valenti}, {Ferraro}, \& {Origlia}}]{val07}
{Valenti}, E., {Ferraro}, F.~R., \& {Origlia}, L. 2007, \aj, 133, 1287

\bibitem[{{Webbink}(1985)}]{web85}
{Webbink}, R.~F. 1985, in IAU Symposium, Vol. 113, Dynamics of Star Clusters,
  ed. J.~{Goodman} \& P.~{Hut}, 541--577

\bibitem[{{Welch} \& {Stetson}(1993)}]{wel93}
{Welch}, D.~L., \& {Stetson}, P.~B. 1993, \aj, 105, 1813

\bibitem[{{Yoon} {et~al.}(2008){Yoon}, {Joo}, {Ree}, {Han}, {Kim}, \&
  {Lee}}]{yoo08}
{Yoon}, S.-J., {Joo}, S.-J., {Ree}, C.~H., {et~al.} 2008, \apj, 677, 1080

\bibitem[{{Yoon} \& {Lee}(2002)}]{yoo02}
{Yoon}, S.-J., \& {Lee}, Y.-W. 2002, Science, 297, 578

\bibitem[{{Zechmeister} \& {K{\"u}rster}(2009)}]{zec09}
{Zechmeister}, M., \& {K{\"u}rster}, M. 2009, \aap, 496, 577

\end{thebibliography}
\end{document}